\documentclass[twocolumn]{aastex631}
\usepackage{gensymb}

\def\Msun{M$_{\odot}$}

\begin{document}

\title{Low-mass stellar and substellar candidate companions around massive stars in Sco OB1 and M17\footnote{Based on observations collected at the European Southern Observatory under ESO programmes 095.D-0495(A), 0101.C-0305(F), 0103.C-0803(A) and 111.24YU.002.}}

\author{Tinne Pauwels}
\affiliation{Institute of Astronomy, KU Leuven, Celestijnenlaan 200D, bus 2401, 3001 Leuven, Belgium}

\author{Maddalena Reggiani}
\affiliation{Institute of Astronomy, KU Leuven, Celestijnenlaan 200D, bus 2401, 3001 Leuven, Belgium}

\author{Hugues Sana}
\affiliation{Institute of Astronomy, KU Leuven, Celestijnenlaan 200D, bus 2401, 3001 Leuven, Belgium}

\author{Laurent Mahy}
\affiliation{Royal Observatory of Belgium, Avenue Circulaire/Ringlaan 3, 1180 Brussels, Belgium}




\begin{abstract}
Massive stars are recognized for their high degree of multiplicity, yet the mass ratio regime below 0.1 remains insufficiently explored. It is therefore unknown whether extremely low-mass (possibly substellar) companions can form and survive in the direct UV-irradiated environment of massive stars. 
In this paper, we discuss VLT/SPHERE IFS (0\farcs15 - 0\farcs85) observations of six massive O- and early B-type stars in Sco OB1 and M17 that each have a low-mass candidate companion. Two targets have companions that are brown dwarf candidates. The other four have candidate companions in the low end of the stellar mass regime ($\leq 0.30$ \Msun). For three of these, we have obtained a second epoch observation. At least two sources exhibit similar proper motion to that of their central star. However, given the expected proper motion of background objects, this does not imply certain companionship. We show how future follow-up observations of the brown dwarf candidate companions in $J$, $H$ and $L$ bands should allow for an unambiguous confirmation of their nature.

\end{abstract}



\section{Introduction} \label{sec:intro}
The vast majority of massive stars are found in binaries and multiple systems \citep{2009Mason, 2014Sana,2017Moe}, but the origin of stellar multiplicity is still uncertain. One possibility is that stellar companions are formed through disk fragmentation, where disks fragment when they become gravitationally unstable \citep{2006Kratter}. For low-mass stars, this mechanism is expected to create companions at relatively close initial separations \citep[10 - 500 AU,][]{2022Offner}.

Observational evidence for protostellar disks around massive stars exists up to masses of at least ~25-30 \Msun\ and sizes of hundreds to a couple of thousand AU \citep{2016Beltran,2015Johnston,2020Goddi}.
Moreover, it has been suggested that massive disks are more likely to be gravitationally unstable than their lower-mass counterparts \citep{2006Kratter}. Because the strong radiation of massive stars stabilizes the disk against fragmentation, the fragmentation happens preferably in the outer regions, with fragmentation typically setting in at $\sim100-200$ AU \citep{2006Kratter}. Simulations of disk fragmentation around massive stars have shown that fragments typically have masses of the order of $\sim 1$ \Msun\ or more massive \citep{2020Oliva}.

Observational constraints on the multiplicity properties of massive stars are of crucial importance in order to understand their formation. More than 70\% of massive stars have a spectroscopic companion and including companions detected through interferometry increases that fraction to almost 100\% \citep{2012Sana,2012Kobulnicky,2013Sana,2014Sana}.
However, the mass ratio regime below 0.1, corresponding to solar- or lower-mass companions, is insufficiently explored as this requires observational techniques that reach extreme contrasts. For spectroscopic binaries, mass ratios close to 0.1 can be reached by using spectral disentangling techniques \citep{2022Shenar, 2022Mahy}.

Recently, high-contrast imaging has been proven to be well-suited to find extremely low mass-ratio companions at intermediate and wide separations. For example, the B-star Exoplanet Abundance STudy \citep[BEAST,][]{2021Janson} investigated the exoplanet frequency around B-type stars through high-contrast imaging observations (VLT/SPHERE) of 85 B-type stars in Sco-Cen. So far, the study has proven to be successful with the discovery of a $10.9 \pm 1.6$ M$_\mathrm{J}$ exoplanet around the 6-10 \Msun\ binary b Centauri \citep{2021Janson_Nature} and two substellar companions ($14.4 \pm 0.8$ M$_\mathrm{J}$ and a probable $18.5 \pm 1.5$ M$_\mathrm{J}$ object) around a 9 \Msun\ B-type star \citep{2022Squicciarini}.

It is unclear how these substellar objects could have been formed and survived around massive stars, since their strong UV radiation creates a harsh environment for such low-mass objects to be formed \citep{2000Armitage, 2019Nicholson}. For this reason, it has been proposed that substellar companions might have been created through capture (free-floating object) or theft (steal a companion from another star). Through N-body simulations, \cite{2022Parker} found that the planets observed by the BEAST survey are potentially captured or stolen: such a scenario occurs on average once in the first 10 Myr of an OB association with density similar to Sco-Cen. The number of companions peaks around an age of 0.1-2 Myr, but many of those are not stable in the long term \citep{2022Daffern-Powell}.

While the formation and survival of brown dwarfs and planets so close to a 9~M$_\odot$ object is already puzzling, the situation is even more perplexing for many low-mass (sub)stellar candidate companions that have been discovered around O-type stars and B-type supergiants \citep{2020Rainot, 2022Rainot, 2021Reggiani, 2023Pauwels}. Indeed, the amount of ionizing radiation increases by three orders of magnitude between an 8 M$_\odot$ B2~V star ($T_\mathrm{eff}\sim22$~kK) and a 16~M$_\odot$ O8.5~V star ($T_\mathrm{eff}\sim32$~kK) \citep{2003Sternberg}. However, it remains uncertain whether these candidate companions are truly bound or whether they are spurious associations that arize due to line-of-sight alignment. Confirming that some of these companions are bound through follow-up observations is challenging because OB stars are typically located in distant star-forming regions and OB associations. Consequently, OB stars exhibit small proper motion and parallax. Despite these difficulties, it is crucial to attempt to confirm that some of the low-mass candidate companions are indeed gravitationally bound to their central massive star.

In this paper, we discuss VLT/SPHERE observations of four stellar low-mass candidate companions (CCs) that have been discovered in the IFS field of view of two O- and two B-type stars in Sco OB1 \citep{2021Reggiani,2023Pauwels}: CPD~$-41\degree$~7721 (O9.7~V), HD152200 (O9.7~IV), HD152042 (B0.5~IV) and HD152385 (B1.5~V). These systems have mass ratios between 0.006 and 0.02, corresponding to estimated companion masses of $0.10-0.30$ \Msun. We obtained VLT/SPHERE follow-up observations for three of these objects (all except CPD~$-41\degree$~7721). In addition, we report the discovery of two potential brown dwarf companions around a B1~Ib-type star in Sco OB1 (HD151805) at $\sim$0\farcs50 ($\sim$765 AU) and one potential brown dwarf companion around an O8.5~V-type star in M17 (Cl* NGC6618 CEN16, hereafter CEN16) at 0\farcs62 ($\sim$1100 AU). These candidate companions were also detected with SPHERE/IFS.

\begin{table*}[!ht]
    \centering
    \caption{Observing conditions. DIT and NDIT listed are for IFS flux (F) and object (O) observations.}
    \begin{tabular}{lccccccccc}
    \hline
    \hline
         Target & Date & DIT (F) & NDIT (F) & DIT (O) & NDIT (O) & Airmass & $\tau_0$ & Seeing & $\Delta$PA \\
         & & [s] & & [s] & & & [s] & [\arcsec] & [\degr] \\
         \hline
         HD~152042 & 2019-06-29 & 32 & 2 & 64 & 10 & 1.05  & 0.0017 & 1.43 & 6.70 \\
         HD~152042 & 2023-05-14 & 32 & 2 & 64 & 12 & 1.05 & 0.0056 & 0.63 & 9.38 \\
         HD~152200 & 2015-08-19 & 16 & 16 & 16 & 10 & 1.29 & 0.0016 & 1.41 & 14.06 \\
         HD~152200 & 2023-05-16 & 8 & 8 & 64 & 12 & 1.08 & 0.0045 & 0.54 & 5.72 \\ 
         HD~152385 & 2019-05-26 & 16 & 4 & 64 & 10  & 1.06 & 0.0022 & 1.14 & 5.83 \\ 
         HD~152385 & 2023-05-16 & 16 & 4 & 64 & 12 & 1.05 & 0.0040 & 0.58 & 9.47 \\ 
         CEN~16\tablenotemark{a} & 2015-08-01 & 16 & 16 & 16 & 5 & 1.07 & 0.0028 & 0.80 & 3.70 \\
         CEN~16 & 2018-08-15 & 2 & 20 & 64 & 30 & 1.01 & 0.0070 & 0.53 & 45.27 \\
         HD~151805 & 2019-05-26 & 16 & 4 & 64 & 10 & 1.05 & 0.0026 & 1.14 & 7.01 \\
         CPD~$-41\degree$~7721 & 2015-08-23 & 16 & 16 & 16 & 10 & 1.14 & 0.0068 & 0.98 & 8.63 \\
         \hline
    \end{tabular}
    \label{tab:observingconditions}
    \tablenotetext{a}{Data quality insufficient.}
\end{table*}

\begin{table*}[!ht]
    \centering
    \caption{Characteristics of target stars}
    \begin{tabular}{lccccc}
    \hline
    \hline
         Target & Spectral Type & $M$ & pmRA$_\mathrm{*}$ & pmDEC$_\mathrm{*}$ & Cluster\\
         & & [\Msun] & [mas/yr] & [mas/yr] & \\
         \hline
         HD~152042 & B0.5 IV$^a$ & 16$^f$ & $-0.42 \pm 0.04^g$ & $-2.05 \pm 0.03^g$ & Sco OB1\\
         HD~152200 & O9.7 IV$^b$ & 17 & $-0.92 \pm 0.03$ & $-1.75 \pm 0.02$ & Sco OB1  \\
         HD~152385 & B1.5 V$^c$ & 10 & $-0.46 \pm 0.05$ & $-1.51 \pm 0.04$ & Sco OB1  \\
         CPD~-41\degree 7721 & O9.7 V$^b$ & 15 & $-0.76 \pm 0.03$ & $-2.00 \pm 0.02$ & Sco OB1  \\
         HD~151805 & B1 Ib$^d$ & 10 & $-0.88 \pm 0.03$ & $-2.15 \pm 0.02$ & Sco OB1  \\
         CEN~16 & O8.5 V$^e$ & 14$^e$ & $0.34 \pm 0.02$ & $-1.60 \pm 0.02$ & M17  \\
         \hline
    \end{tabular}
    \label{tab:targetchar}\\
    
    \footnotesize{Notes. $^a$\cite{1977Garrison}, $^b$\cite{2014Sota}, $^c$R. H. Barbá and J. Maíz
Apellániz: private communication, $^d$\cite{1978Houk}, $^e$\cite{2017RamirezTannus}, $^f$Masses for targets in Sco OB1 were estimated from the Hertzsprung-Russell diagram in \cite{2023Pauwels}, $^g$Proper motion from \cite{2023Gaia,2016Gaia}.}
\end{table*}

\section{Observations and data reduction}
The observations were taken with the Spectro-Polarimetric High-contrast Exoplanet REsearch instrument \citep[SPHERE,][]{2019Beuzit} at the Very Large Telescope (VLT). Table \ref{tab:observingconditions} shows the observing setup, atmospheric conditions and parallactic angle variation ($\Delta$PA). $\tau_0$ is the average coherence time during each observation. The targets were observed between 2015 and 2023. Four out of six targets have a first and second epoch observation, while two targets (HD~1515805 and CPD~$-41\degree$~7721) only have a first epoch observation. However, the quality of the first epoch of CEN16 (2015) is not sufficient to characterize its low-mass companion, so we only consider the second epoch (2018). The time between two epochs varies per target: $\sim 3.9$ yrs for HD152042, $\sim 7.7$ yrs for HD152200, and $\sim 4.0$ yrs for HD152385. The spectral type, mass ($M$), proper motion (pmRA$_\mathrm{*}$, pmDEC$_\mathrm{*}$) and cluster environment of the targets is shown in Table \ref{tab:targetchar}.

The observations were executed in IRDIFS\_EXT mode, using two SPHERE science sub-systems: the Integral Field Spectrograph \citep[IFS,][]{2008Claudi, 2015Mesa} and the Infra-Red Dual-band Imager and Spectrograph \citep[IRDIS,][]{2008Dohlen,2010Vigan}. The first instrument (IFS) observes in a field of view (FoV) of 1.73\arcsec x 1.73\arcsec{} and in 39 wavelength bands ($0.95-1.65$ $\mu$m), allowing us to extract a low-resolution spectrum for potential companions and characterize their fundamental properties. The latter observes in a larger portion of the sky of 11\arcsec x 12.5\arcsec{} and in only two wavelength bands ($K_1=2.11 $ $\mu$m and $K_2 = 2.25$ $\mu$m). All companions discussed in this paper are observed in the IFS FoV, so that we can extract their spectrum. Evidently, these sources are also in the IRDIS field, so that we obtain a total of 41 wavelength bands. 

The observing sequence consisted of \textsc{flux}, \textsc{center} and \textsc{object} observations. First, the flux (F) observations are off-axis observations of the central star, allowing to obtain a point spread function (PSF) that is used to calibrate the companion spectrum. A neutral density filter (ND2.0) was added to avoid saturation.
Second, center (C) frames are used to center the coronagraph on the central star. Thirdly, object (O) observations are taken with a coronagraph that blocks the light from the central star to increase the contrast at close separations.

The data reduction was performed by the High Contrast Data Center (HC-DC, previously SPHERE Data Center) \citep{2017Delorme,2018Galicher}. They handle the centering of the central star in the IRDIS and IFS images and perform the astrometric calibration of the images. We further adopt a platescale of $7.46 \pm 0.02$ mas/pixel \citep{2016Maire} for IFS and $12.258 \pm 0.004$ mas/pixel for IRDIS \citep{2021Maire}.

\begin{figure*}
\gridline{\fig{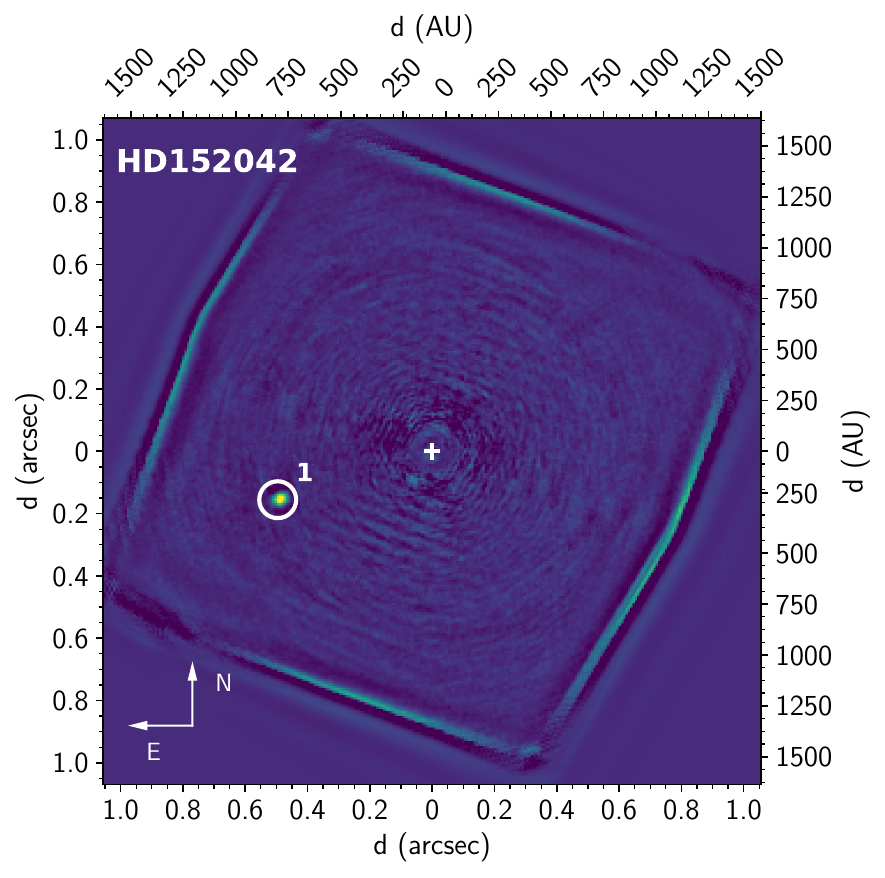}{0.33\textwidth}{(a)}
          \fig{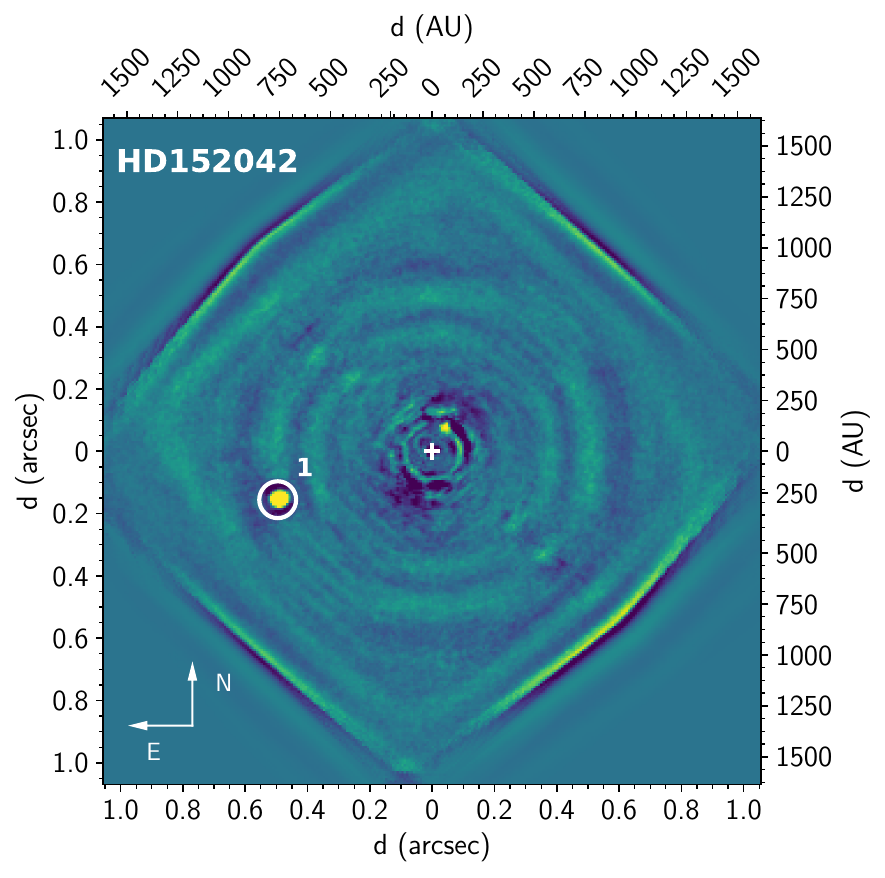}{0.33\textwidth}{(b)}
          \fig{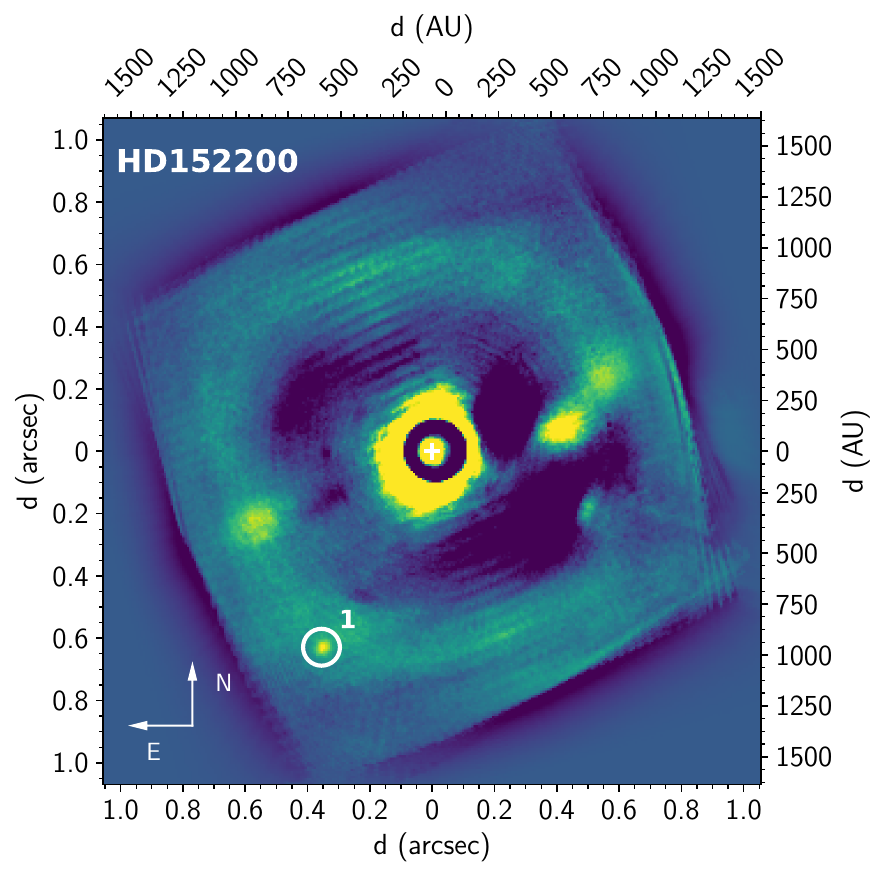}{0.33\textwidth}{(c)}}
\gridline{\fig{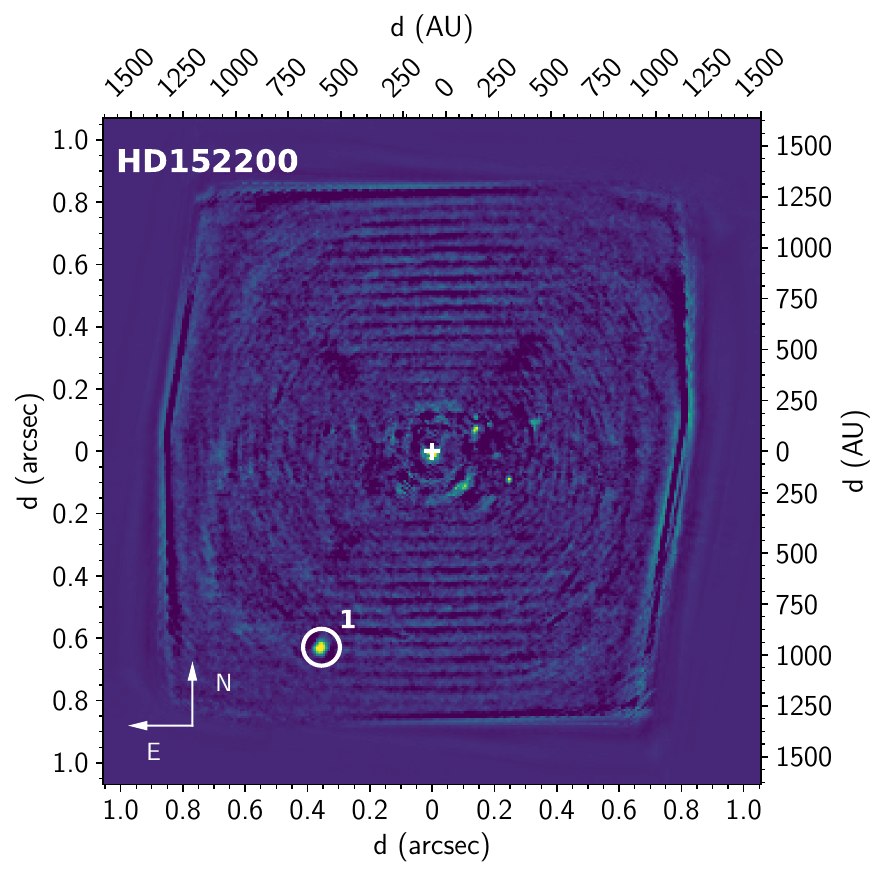}{0.33\textwidth}{(d)}
          \fig{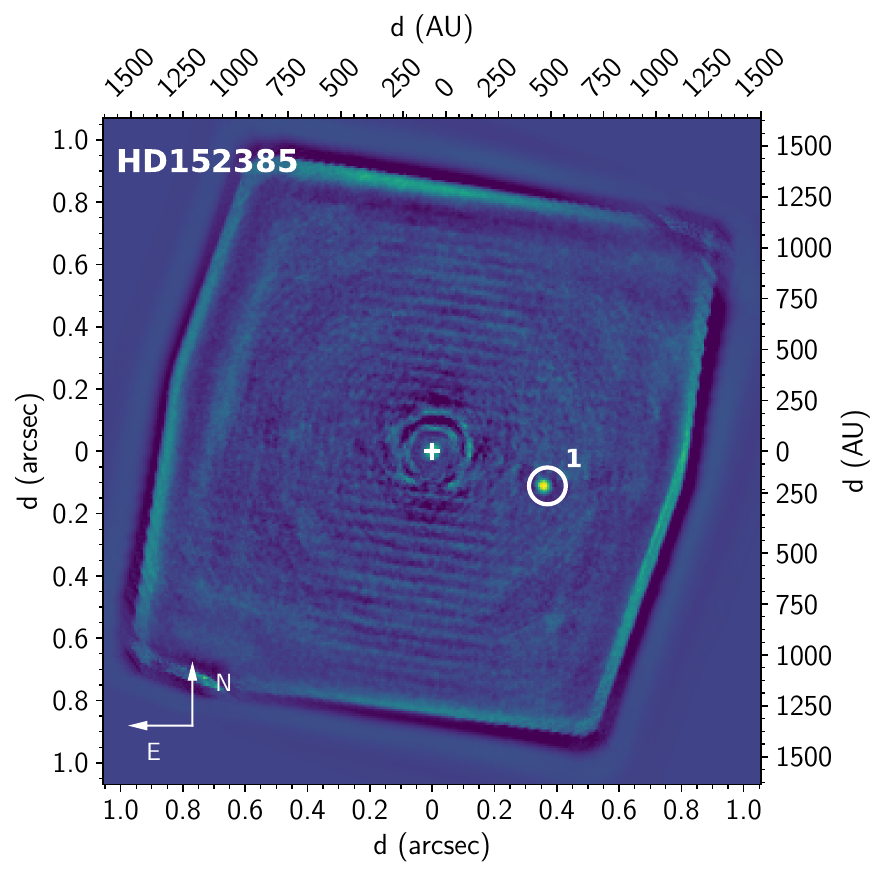}{0.33\textwidth}{(e)}
        \fig{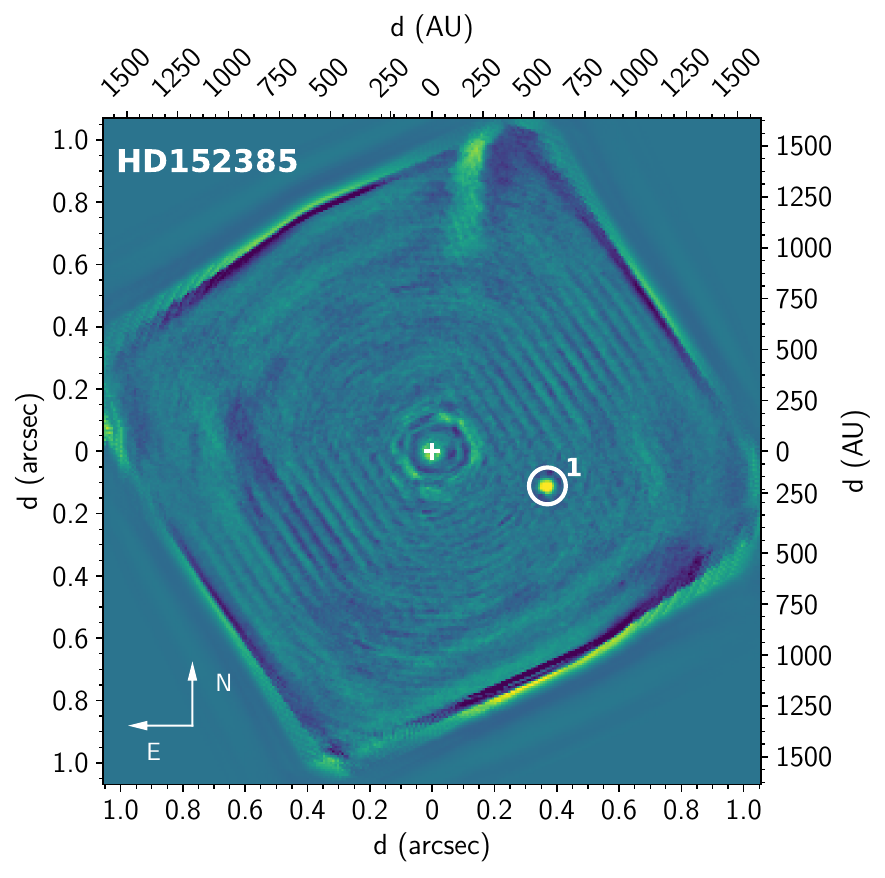}{0.33\textwidth}{(f)}}
\gridline{\fig{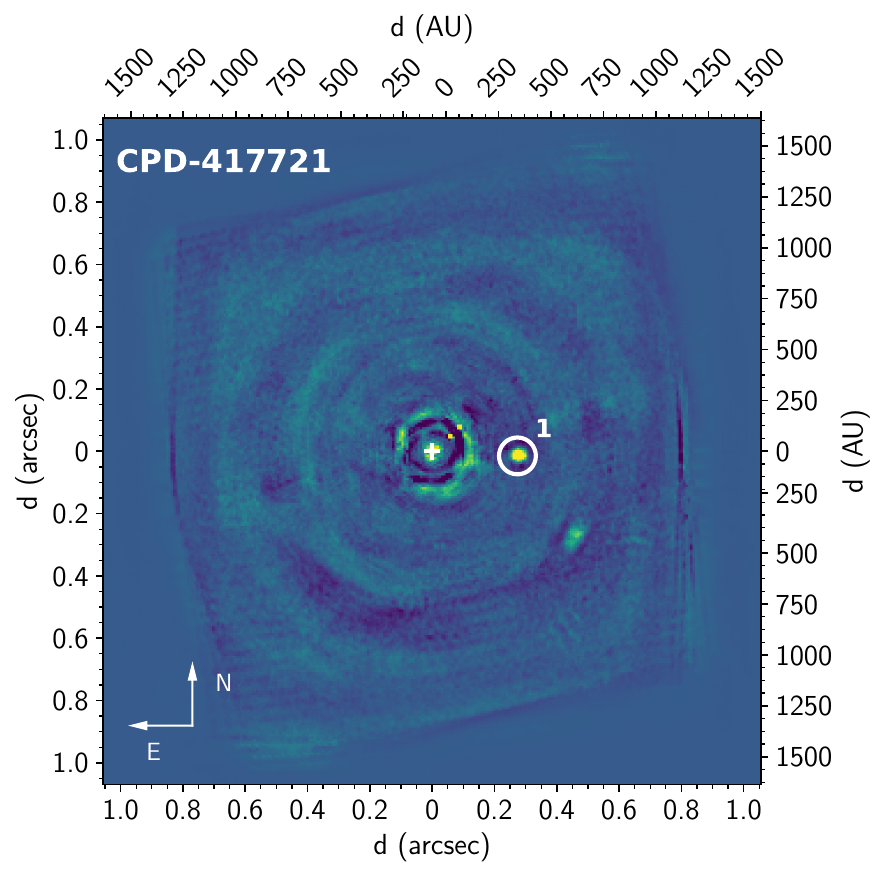}{0.33\textwidth}{(g)}}
\caption{PCA/SDI IFS images of stars with a stellar candidate companion. (a) First epoch of HD~152042 with eight PCs. (b) Second epoch of HD~152042 with five PCs. (c) First epoch of HD~152200 with 23 PCs. (d) Second epoch of HD~152200 with 23 PCs. (e) First epoch of HD~152385 with three PCs. (f) Second epoch of HD~152385 with five PCs. (g) Only epoch of CPD~-41\degree 7721 with 11 PCs.}
\label{fig:SDIstellar}
\end{figure*}

\begin{figure*}[!t]
\gridline{\fig{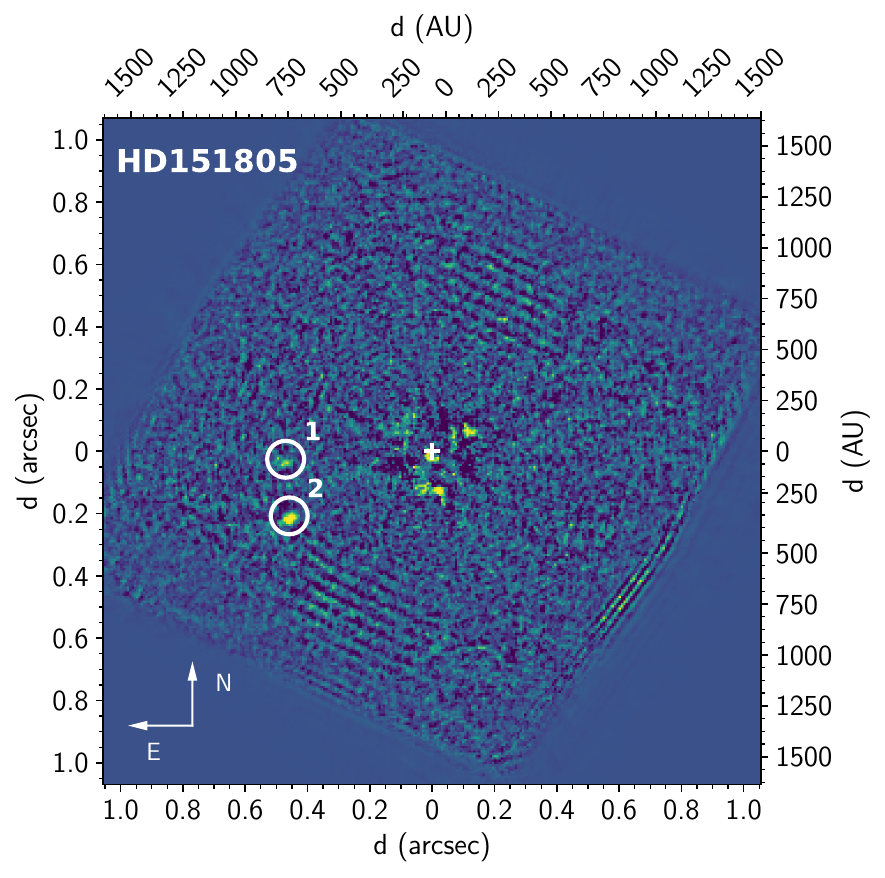}{0.33\textwidth}{(a)}
          \fig{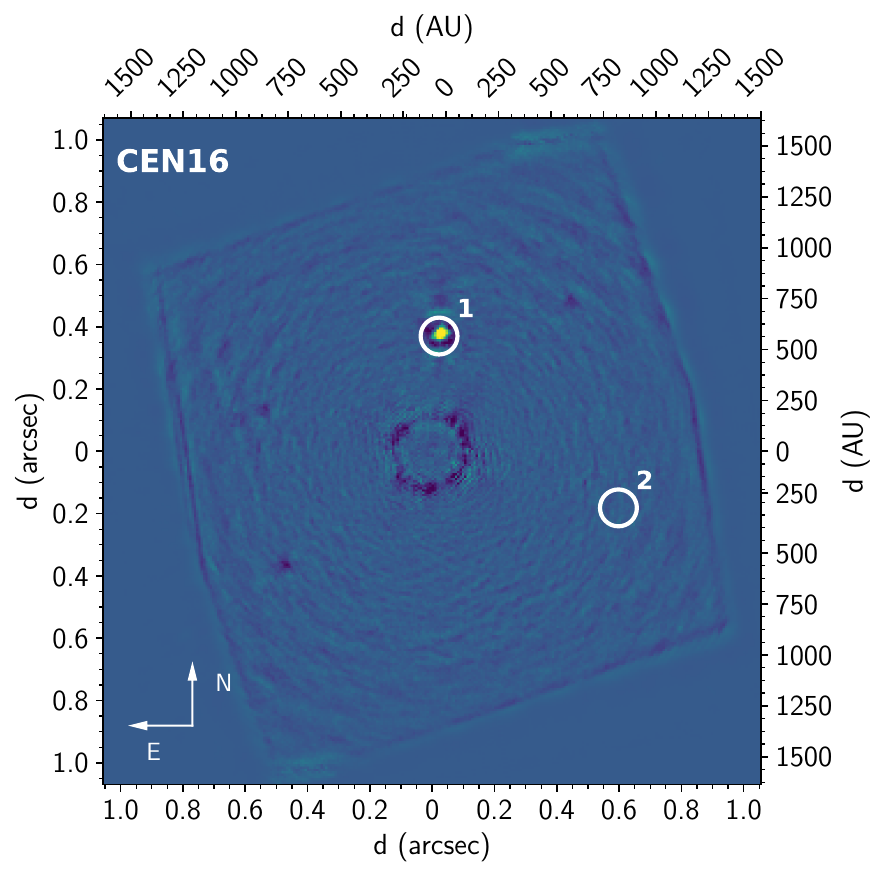}{0.33\textwidth}{(b)}
          \fig{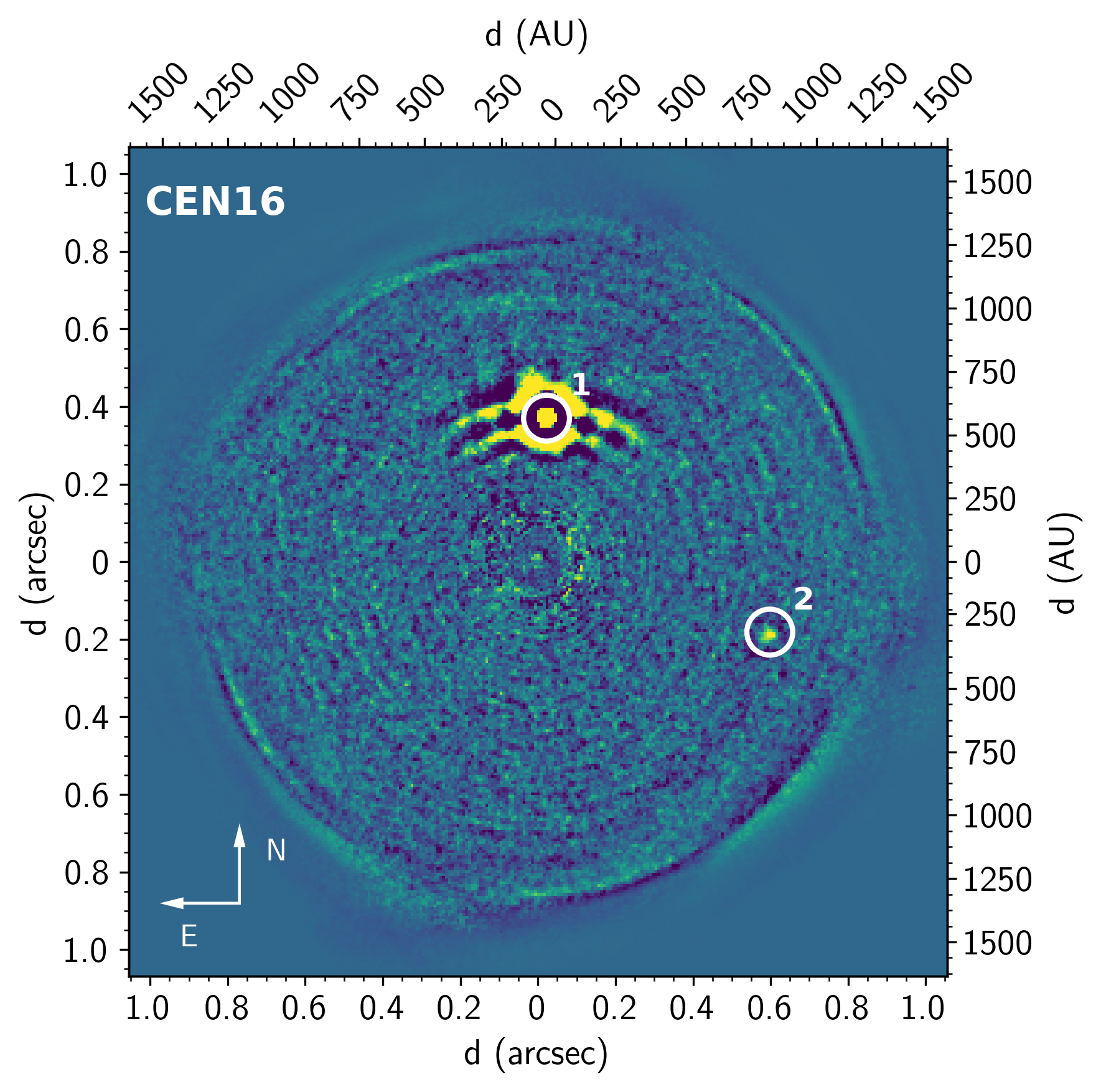}{0.33\textwidth}{(c)}}
\caption{PCA/SDI IFS images of stars with (a) substellar candidate companion(s). (a) Only epoch of HD~151805 with 26 PCs. (b) First epoch of CEN16 with 40 PCs. (c) Second epoch of CEN16 with 58 PCs.}
\label{fig:SDIsubstellar}
\end{figure*}

\section{Data analysis and results}
We mainly follow the method described in \cite{2020Rainot, 2022Rainot}, \cite{2021Reggiani} and \cite{2023Pauwels}. We use the Vortex Image Processing (VIP) Python package to perform the image processing and data analysis \citep[v.1.4.0,][]{2017GomezGonzalez,2023Christiaens}. The PCA/SDI IFS images are shown in Fig.\ref{fig:SDIstellar} for stellar candidate companions and Fig.\ref{fig:SDIsubstellar} for substellar candidate companions. The number of principal components (PCs) was chosen to maximize the signal-to-noise ratio (S/N). The candidate companions are labeled with a number. If more than one candidate companion is detected in the IFS FoV, they are labeled in ascending order, with `1' for the closest and `2' for the next closest to the central star. We will refer to them as `CC1' and `CC2'.

\subsection{Astrometry retrieval}
We follow the method proposed by \cite{2017Wertz} to obtain robust astrometry of the candidate companions. We perform angular differential imaging (ADI) on the IFS images and calculate the number of PCs that leads to the highest S/N value of the candidate companion for each spectral channel. To speed up the process, the next calculations are performed using only the eight channels with the highest S/N. A first guess of the position is obtained by identifying the highest pixel value in the neighborhood of the companion. The position is fixed and a negative fake companion \citep[NEGFC,][]{2010Lagrange,2010Marois} is then injected to estimate the flux. The NEGFC technique consists of injecting a negative PSF template into the cube and trying to cancel out the companion in the post-processed ADI image, using the ideal number of PCs that was calculated before. Finally, a Nelder-Mead optimization refines the position ($r$, $\theta$) to sub-pixel precision and adjusts the flux accordingly. Table \ref{tab:astrometry} presents the astrometric results.

\subsubsection{Sources of astrometric uncertainties}
As explained in detail in \cite{2017Wertz}, there are four main sources that contribute to the astrometric uncertainties: (i) the instrumental calibration errors, (ii) the centering error of the central star, (iii) the statistical error related to the determination of the companion position with respect to the central star and (iv) the systematic error due to residual speckles.

We adopt the astrometric calibration values determined by \cite{2021Maire}, which were derived from five years of monitoring IRDIS data. The sources of instrumental calibration uncertainties include the platescale, the True North orientation angle of $-1.76 \pm 0.04$\degree, the pupil offset of $136.00 \pm 0.03$\degree{}, the dithering uncertainty of $0.74$ mas (for IRDIS observations only) and the optical distortion (pixel scale ratio between the horizontal and vertical directions of the IRDIS detector) of $1.0075 \pm 0.0004$. We include the optical distortion as an additional uncertainty of $0.04\%$ on the platescale. Since the IFS detector is offset by $-100.48 \pm 0.13$\degree{} relative to the IRDIS detector, we account for this additional uncertainty for IFS observations \citep{2016Zurlo}.

We assume a conservative centering error of the central star of 0.2 pixels, translating to 1.5 mas for IFS and 2.5 mas for IRDIS \citep{2018Chauvin}.

Finally, the statistical error is obtained through bayesian inference with Markov Chain Monte Carlo (MCMC) sampling with 100 walkers. The convergence is evaluated based on the auto-correlation time $\tau$. We assume that convergence is reached when $\tau/N < 0.02$, with $N$ the number of steps. Typically, between 5000-6000 steps are needed to reach convergence. For the targets that only have a first epoch observations (CPD~417721, HD~151805 and CEN~16), we limit the MCMC sampling to 5000 and visually check that the chains are sufficiently converged. This is the case for all except HD~151805 CC1, where the non-convergence is evident from the large errors on the astrometry (Table \ref{tab:astrometry}). This is likely due to the very low flux detected from this companion, which hinders effective MCMC sampling. Since the error calculated through MCMC sampling also includes the effect of residual speckles, we do not calculate a separate systematic uncertainty.

\begin{table*}[!t]
    \centering
    \caption{Characteristics of candidate companions. The position angles $\theta$ are measured north to east. $\Delta K_1$ and $\Delta K_2$ are the contrast magnitudes in the IRDIS $K_1$ and $K_2$ bands.}
    \begin{tabular}{lccccccc}
    \hline
    \hline
         Target & Date & $r_\mathrm{IFS}$ & $\theta_\mathrm{IFS}$ & $r_\mathrm{IRDIS}$ &$\theta_\mathrm{IRDIS}$ & $\Delta K_{1}$ & $\Delta K_{2}$\\
         & & [mas] & [\degr] & [mas] & [\degr] & [mag] & [mag] \\
         \hline
         HD~152042 CC1 & 2019-06-29 & $516.6 \pm 1.9$ & $107.25 \pm 0.18$ & $520.9 \pm 1.3$ & $107.56 \pm 0.15$ & $7.66 \pm 0.04$ & $7.44 \pm 0.07$ \\
         HD~152042 CC1 & 2023-05-14 & $518.3 \pm 2.0$ & $107.15 \pm 0.18$ & $522.3 \pm 1.4$ & $107.59 \pm 0.15$ & $7.54 \pm 0.04$ & $7.34 \pm 0.07$\\
         HD~152200 CC1 & 2015-08-19 & $725.7 \pm 4.0$ & $150.89 \pm 0.22$ & $729.7 \pm 2.7$ & $150.48 \pm 0.19$ & $8.74 \pm 0.14$ & $8.52 \pm 0.16$\\
         HD~152200 CC1 & 2023-05-16 & $731.9 \pm 2.5$ & $150.30 \pm 0.18$ & $737.5 \pm 2.9$ & $150.37 \pm 0.18$ & $8.87 \pm 0.10$ & $8.59 \pm 0.15$ \\ 
         HD~152385 CC1 & 2019-05-26 & $374.4 \pm 2.9$ & $252.46 \pm 0.27$ & $373.0 \pm 3.2$ & $252.09 \pm 0.27$ & $7.58 \pm 0.10$ & $7.68 \pm 0.24$\\ 
         HD~152385 CC1 & 2023-05-16 & $384.5 \pm 2.4$ & $252.83 \pm 0.25$ & $382.6 \pm 3.5$ & $252.56 \pm 0.28$ & $7.56 \pm 0.10$ & $7.57 \pm 0.24$\\ 
         CPD~$-41\degree$~7721 CC1 & 2015-08-23 & $277.9 \pm 2.3$ & $267.35 \pm 0.38$ & $282.7 \pm 4.1$ &  $266.65 \pm 0.63$ & $7.42 \pm 0.11$ & $7.33 \pm 0.21$ \\
         CEN~16 CC2 & 2018-08-15 & $623.0 \pm 5.6$ & $252.13 \pm 0.43$  & $628.1 \pm 6.0$ & $251.49 \pm 0.41$ & $8.70 \pm 0.24$ & $8.59 \pm 0.18$\\
         HD~151805 CC1 & 2019-05-26 & $497.4 \pm 19.3$ & $94.79 \pm 1.54$ & $495.2 \pm 32.3$ & $94.22 \pm 3.57$ & $9.88 \pm 0.79$ & $11.46 \pm 0.26$ \\
         HD~151805 CC2 & 2019-05-26 & $512.2 \pm 5.6$ & $115.07 \pm 0.40$  & $514.8 \pm 4.9$ & $115.56 \pm 0.40$ &  $9.26 \pm 0.15$ & $9.74 \pm 0.27$  \\
         \hline
    \end{tabular}
    \label{tab:astrometry}
\end{table*}

\subsubsection{Final astrometric errors}
The final astrometric uncertainties are calculated as follows:
\begin{equation}
           \sigma_{r,\mathrm{tot}}^{2} = \mathrm{PLSC}^{2}\cdot \left(\sigma_{r,\mathrm{MCMC}}^{2} + \sigma_{r,\mathrm{cen}}^{2} + (\sigma_{r,\mathrm{dit}}^{2}) \right) + \sigma_\mathrm{PLSC}^{2} \cdot r^{2} 
\end{equation}
and
\begin{equation}
    \sigma_{\theta,\mathrm{tot}}^{2} = \sigma_{\theta,\mathrm{MCMC}}^{2} + \sigma_{\theta,\mathrm{cen}}^{2} +  \sigma_{\mathrm{PO}}^{2} + \sigma_{\mathrm{TN}}^{2} 
    (+ \sigma_{\theta,\mathrm{dit}}^{2}) (+ \sigma_{\mathrm{IFSoffset}}^{2})
\end{equation}
where the $\sigma_{r,\mathrm{MCMC}}$ and $\sigma_{\theta,\mathrm{MCMC}}$ refer to the statistical errors determined with MCMC sampling, $\sigma_{r,\mathrm{cen}}$ and $\sigma_{\theta,\mathrm{cen}}$ are the centering uncertainties in $r$ and $\theta$ directions, $\sigma_{r,\mathrm{dit}}$ and $\sigma_{\theta,\mathrm{dit}}$ are the dithering uncertainties (only included for IRDIS observations), PLSC and $\sigma_\mathrm{PLSC}$ are the platescale and platescale uncertainty, $\sigma_{\mathrm{PO}}$ is the pupil offset error, $\sigma_{\mathrm{TN}}$ is the True North error, and $\sigma_{\mathrm{IFSoffset}}$ is the IFS offset error (only included for IFS observations).

\subsection{Flux extraction and calibration}
Once the position of the sources in the IFS and IRDIS images has been determined, another iteration of the NEGFC technique is performed, during which the position remains fixed while only the flux is allowed to vary. Next, we apply MCMC sampling on a channel-by-channel basis to further optimize the flux and determine the uncertainties, using the same setup as for the astrometry retrieval.

To flux calibrate the companion spectrum, we computed the spectral energy distribution (SED) of the central star with the atmosphere code FASTWIND \citep[v.10.3.2,][]{2005Puls, 2011Rivero, 2018Sundqvist}, using an effective temperature, radius and surface gravity that are based on measured parameters or the spectral type of the central star obtained from literature (see \cite{2017RamirezTannus} for CEN16 and \cite{2023Pauwels} for others). The formulas suggested by \cite{2001Vink} were used to calculate the mass-loss rate and terminal wind velocity. We scaled the FASTWIND model to match the observed $K_s$-band magnitude of the central star \citep{2003Cutri}, accounting for individual distances and K-band extinctions as described in \cite{2023Pauwels}. To verify our method, we performed an atmosphere analysis of an observed FEROS spectrum of HD~151805 with CMFGEN \citep{1998Hillier}. The derived temperature and $\log g$ are $27000 \pm 2000$ K and $3.75\pm 0.25$, leading to an SED that differs by more than a factor 10 in $K$-band from the initial FASTWIND SED. However, after scaling the FASTWIND model, the predicted $K$-band fluxes by FASTWIND and CMFGEN  agree within 20\%. 
We conclude that the physical parameters used to compute the NIR absolute flux of the central star have a negligible impact on the flux calibrated spectrum of the companions, as long as the central-star model is scaled to the observed $K$-magnitude. This is indeed expected as, in the OB star temperature regime, the NIR domain is well within the Rayleigh-Jeans tail of the energy distribution.
For HD151805, we use the CMFGEN SED to flux calibrate the companion spectrum. For CEN16, the FASTWIND model is not scaled to the observed $K$-band magnitude, as the properties of the central star are calculated by \cite{2017RamirezTannus} in a way that already takes into account the observed $K$-band magnitude. To calibrate the flux, the star needs to be positioned at an artificial reference distance. We arbitrarily chose 100 R$_\odot$ for this purpose, but this selection does not impact the results of the analysis.

\begin{figure*}
\gridline{\fig{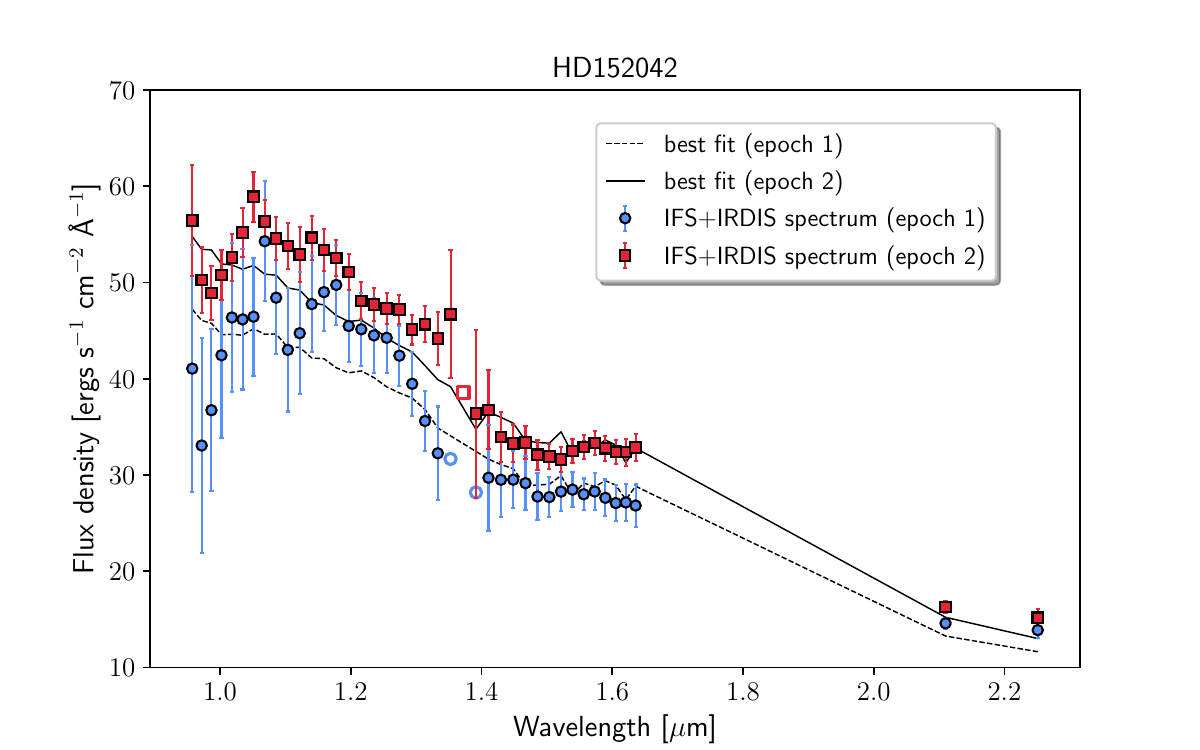}{0.5\textwidth}{(a) HD~152042 CC1}
\fig{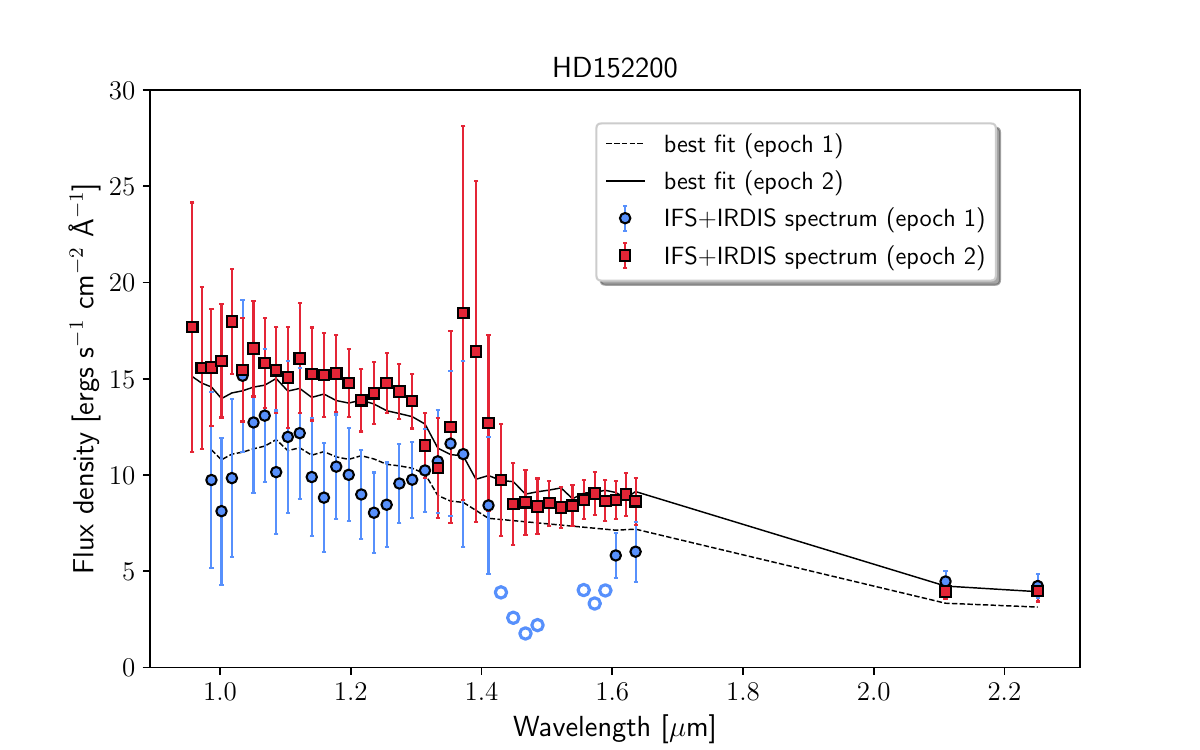}{0.5\textwidth}{(b) HD~152200 CC1}}

\gridline{\fig{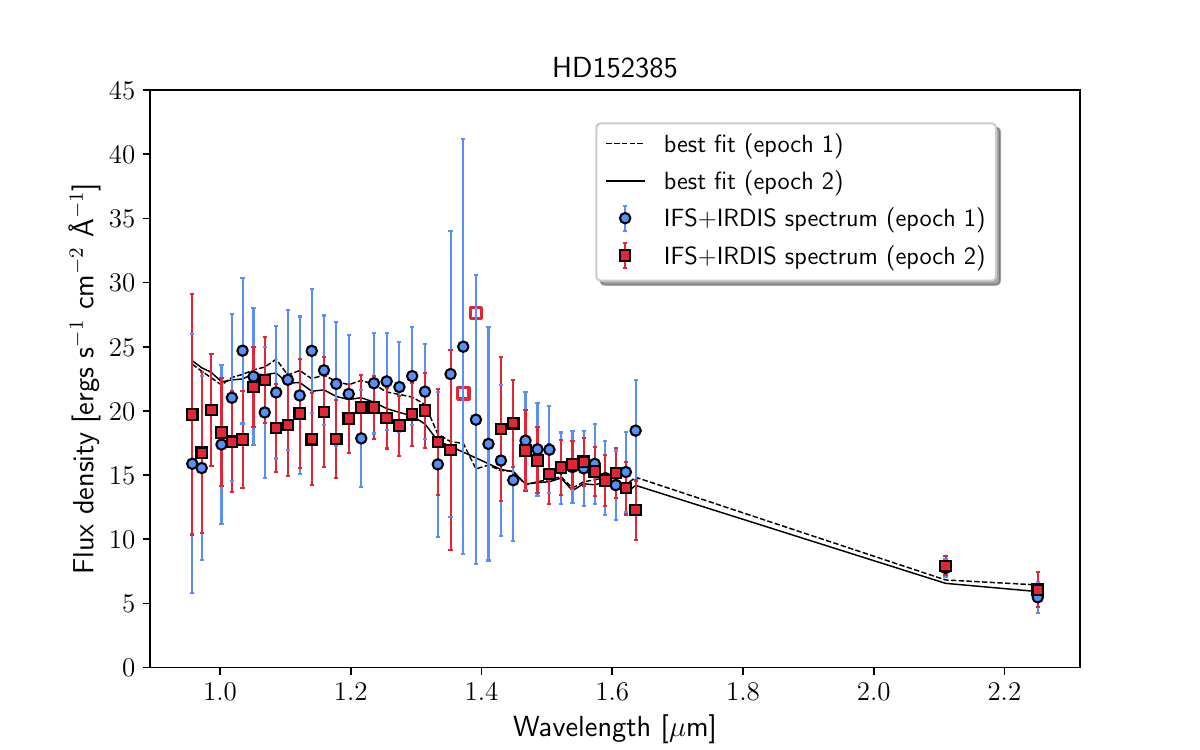}{0.5\textwidth}{(c) HD~152385 CC1}
          \fig{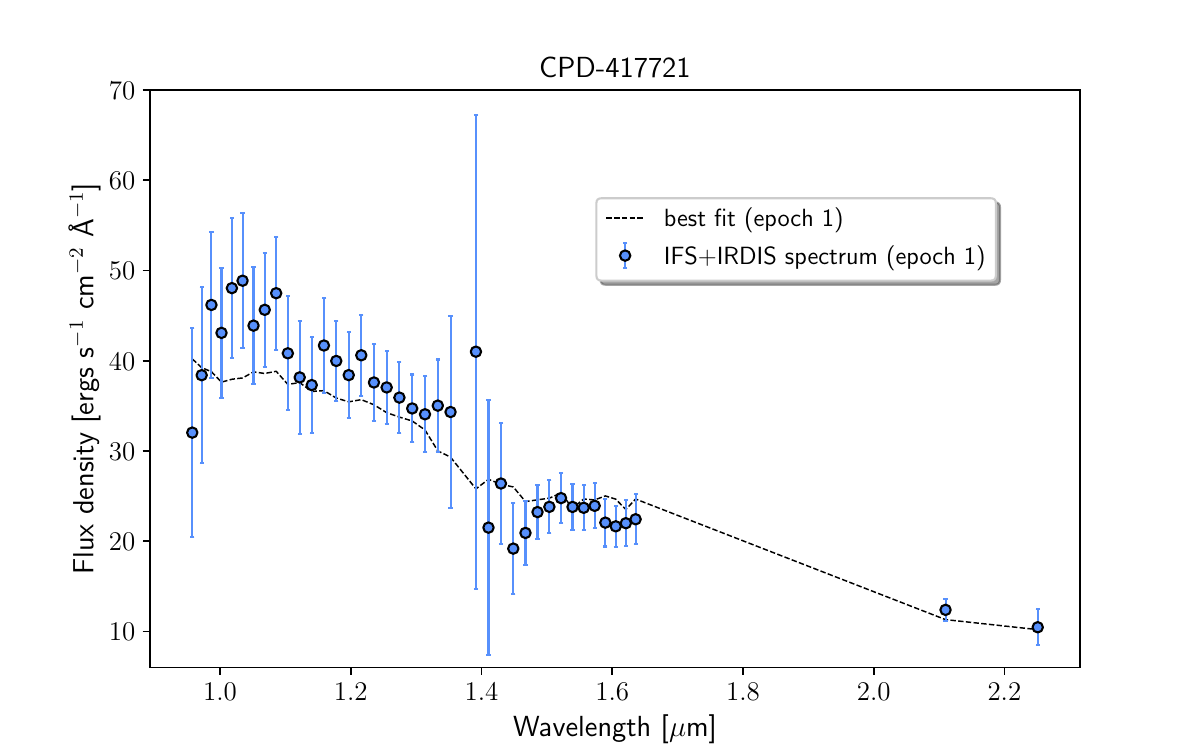}{0.5\textwidth}{(d) CPD-417721 CC1}}
\caption{IFS+IRDIS spectra of stellar candidate companions. The flux density values are plotted with error bars in blue (first epoch) or red (second epoch). The least squares best-fit model is shown with a dashed line (first epoch) or a full line (second epoch). Open symbols represent flux density values that were left out of the least squares fit.}
\label{fig:IFSspecstellar}
\end{figure*}

\begin{figure*}
\gridline{\fig{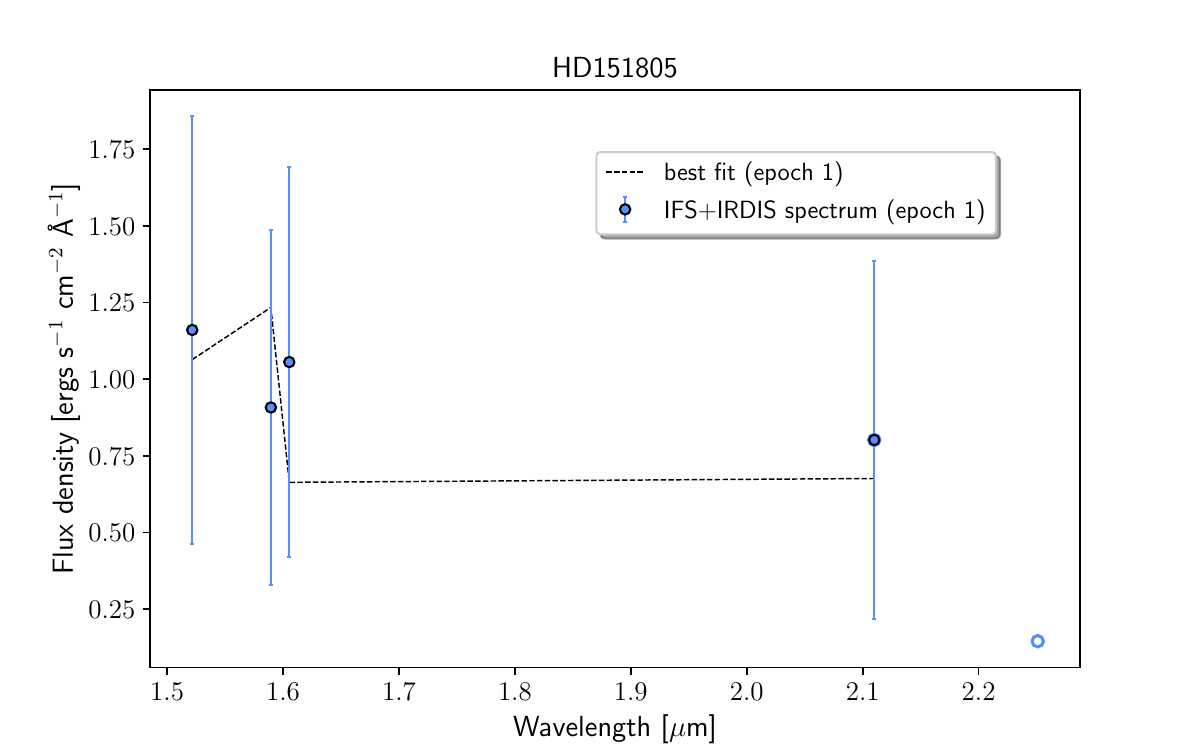}{0.5\textwidth}{(a) HD~151805 CC1}
\fig{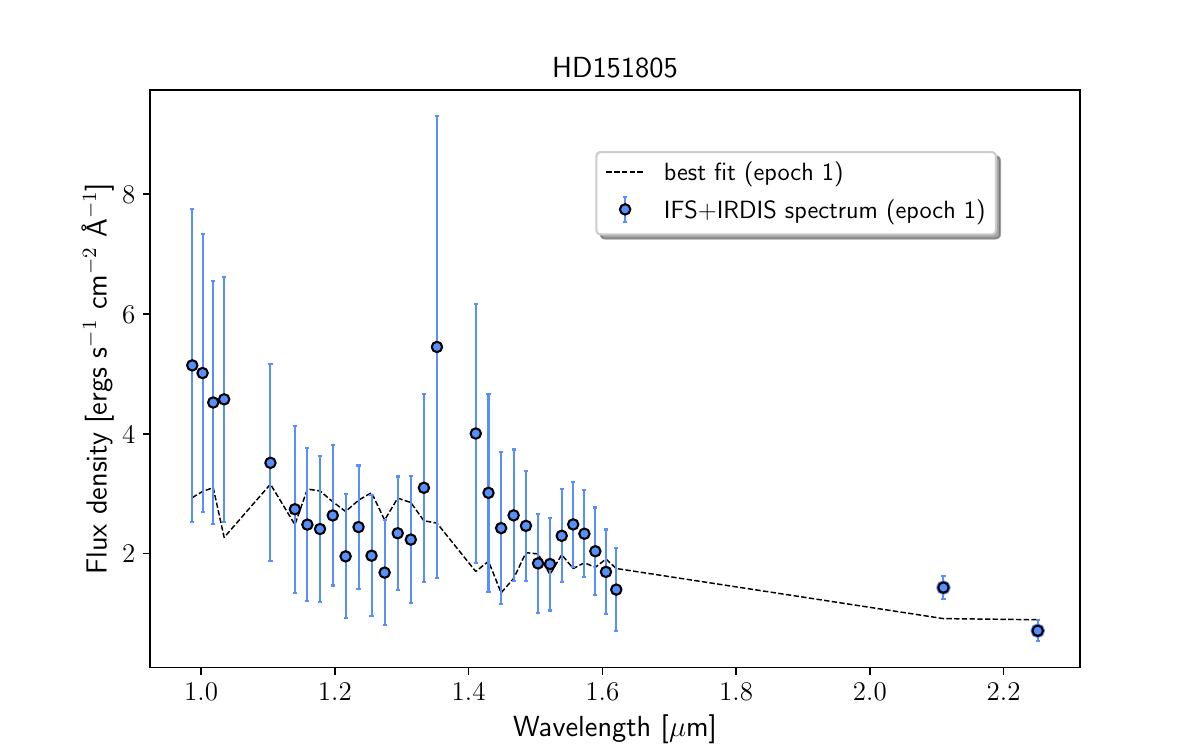}{0.5\textwidth}{(b) HD~151805 CC2}}
\gridline{\fig{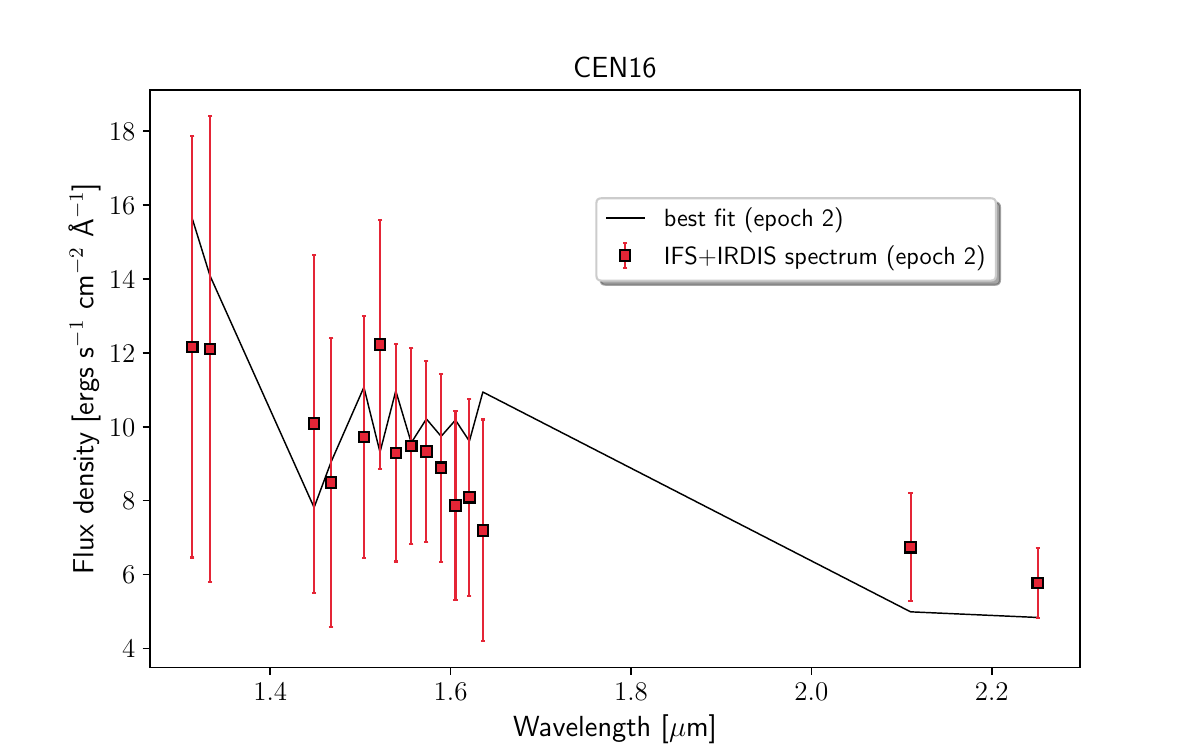}{0.5\textwidth}{(c) CEN16 CC2}}
\caption{IFS+IRDIS spectra of substellar candidate companions. The flux density values are plotted with error bars in blue or red. The least squares best-fit model is shown with a dashed or a full line.} 
\label{fig:IFSspecsubstellar}
\end{figure*}

Finally, we compared the flux calibrated companion spectrum with a family of pre-main sequence atmosphere and evolutionary models \citep{2000Siess,2013Husser,2014Allard} to obtain an estimate of the companions' fundamental parameters by means of a least squares fit. We interpolated the models to match the resolution of the IFS spectra. Since there is a degeneracy between the age and the other physical parameters of the companions, we constrained the fit between 4 and 8 Myrs for stars in Sco OB1 \citep{2013Sung, 2016Damiani} and $\leq 1$ Myr for M17 \citep{2017RamirezTannus}. The spectra and best fits are shown in Fig.\ref{fig:IFSspecstellar} for stellar mass objects and Fig.\ref{fig:IFSspecsubstellar} for substellar mass objects. The error bars on the spectrum are the combination of the errors on the contrast flux of the companions computed with MCMC sampling and the measured flux of the central star, derived by calculating the standard deviation of the off-axis observations of the central star at different parallactic angles. The pixel intensities in calibrated SPHERE images from the HC-DC are normalized by integration time, making it impossible to account for photon noise uncertainty \citep{2023Christiaens}.

The best least squares fit values are presented in Table \ref{tab:IFSfit}. The error bars are obtained by including all the models that fit to the spectrum within $3\sigma$.

\begin{table*}[!ht]
    \caption{IFS+IRDIS spectra least squares fit results}
    \centering
    \begin{tabular}{lccccccc}
    \hline
    \hline
         Target & Epoch & $M$ & $\log T$ & $R$ & $\log L/\mathrm{L}_\odot$ & age & $\chi^2$ \\
         & & [\Msun] & [K] & [R$_\odot$] &  & [Myr] &  \\
         \hline
         HD~152042 CC1 & 1 & $0.25^{+0.15}_{-0.00}$ & $3.52^{+0.02}_{-0.00}$ & $0.92^{+0.04}_{-0.04}$ & $-0.97^{+0.04}_{-0.05}$ & $4.6^{+3.4}_{-0.4}$ & $40.28$ \\
         HD~152042 CC1 & 2 & $0.30$ & $3.53$ & $0.94$ & $-0.90$ & $4.8$ & $103.73$ \\
         HD~152200 CC1 & 1 & $0.10^{+0.05}_{-0.03}$ & $3.48^{+0.02}_{-0.02}$ & $0.55^{+0.10}_{-0.08}$ & $-1.54^{+0.07}_{-0.23}$ & $6.4^{+1.6}_{-2.4}$ & $17.77$ \\
         HD~152200 CC1 & 2 & $0.13^{+0.04}_{-0.06}$ & $3.49^{+0.02}_{-0.03}$ & $0.60^{+0.10}_{-0.05}$ & $-1.43^{+0.07}_{-0.18}$ & $7.0^{+1.0}_{-3.0}$ & $12.79$ \\
         HD~152385 CC1 & 1 & $0.13^{+0.12}_{-0.03}$ & $3.49^{+0.04}_{-0.02}$ & $0.78^{+0.10}_{-0.16}$ & $-1.23^{+0.10}_{-0.20}$ & $5.0^{+3.0}_{-1.0}$ & $11.86$\\ 
         HD~152385 CC1 & 2 & $0.20^{+0.05}_{-0.10}$ & $3.51^{+0.01}_{-0.04}$ & $0.69^{+0.16}_{-0.05}$ & $-1.23^{+0.04}_{-0.20}$ & $6.9^{+1.1}_{-2.9}$ & $21.29$\\
         CPD~$-41\degree$~7721 CC1 & 1 &  $0.20^{+0.15}_{-0.17}$ & $3.51^{+0.03}_{-0.03}$ & $0.90^{+0.07}_{-0.11}$ & $-1.04^{+0.05}_{-0.09}$ & $4.1^{+3.9}_{-0.1}$ & $24.39$ \\
         CEN~16 CC2 & 2 & $0.07^{+0.01}_{-0.02}$ & $3.46^{+0.00}_{-0.02}$ &  $0.75^{+0.09}_{-0.08}$ & $-1.48^{+0.09}_{-0.16}$ & $1.0^{+0.0}_{-0.0}$ & $7.31$  \\
         HD~151805 CC1 & 1 & $0.03^{+0.01}_{-0.01}$ & $3.41^{+0.03}_{-0.02}$ &  $0.32^{+0.08}_{-0.07}$ & $-2.40^{+0.27}_{-0.29}$
         & $7.0^{+1.0}_{-3.0}$ & $0.76$ \\
         HD~151805 CC2 & 1 & $0.04^{+0.02}_{-0.02}$ & $3.44^{+0.02}_{-0.04}$ & $0.33^{+0.07}_{-0.02}$ & $-2.24^{+0.14}_{-0.25}$ & $7.0^{+1.0}_{-3.0}$ & $21.72$ \\
         \hline
    \end{tabular}
    \label{tab:IFSfit}
\end{table*}

\subsubsection{Low-mass stellar (candidate) companions}
For some spectra, certain wavelength channels were left out of the least squares fit because of large error bars or low S/N ($\lesssim 2$). The flux density values of these channels are plotted with open symbols in Fig.\ref{fig:IFSspecstellar} without error bars to improve the clarity of the plot.

HD152042 was observed with SPHERE at two different epochs. The flux measured in the first epoch is lower than that of the second epoch, but still within the error bars. This is likely due to the bad seeing conditions of the first epoch (Table \ref{tab:observingconditions}). We find a mass of $0.25^{+0.15}_{-0.00}$ \Msun\ for the first epoch and $0.30$ \Msun\ for the second epoch. The second epoch values have no error bars because no fit was found within $3\sigma$ due to the density of our grid.

HD152200 was observed with SPHERE at two different epochs. Again, the flux measured in the first epoch is lower than that of the second epoch, likely due to the bad seeing conditions during the first epoch observation, but it is still within the error bars (Table \ref{tab:observingconditions}). For the least squares fit of the first epoch, we left out a dip in the flux between 1.4 and 1.6 $\mu$m, which is not seen in the second epoch and probably a result of the bad seeing. We find a mass of $0.10^{+0.05}_{-0.03}$ \Msun\ for the first epoch and $0.13^{+0.04}_{-0.06}$ \Msun\ for the second epoch.

HD152385 was observed with SPHERE at two different epochs. The measured flux densities of both epochs agree within the error bars. We find a mass of $0.13^{+0.12}_{-0.03}$ \Msun\ for the first epoch and $0.20^{+0.05}_{-0.10}$ \Msun\ for the second epoch.

CPD~-41\degree 7721 was observed once with SPHERE. We find a mass of $0.20^{+0.15}_{-0.17}$ \Msun.

Overall, the results from different epochs, when available, provide consistent mass estimates (within errors).

\subsubsection{Brown dwarf candidate companions}
The brown dwarf candidates are bright enough to extract their flux for some of the redder IFS  channels, but are too faint at shorter wavelengths. We only extracted the flux of channels that have S/N$>2$. The extracted IFS spectra and $K_1$- and $K_2$-band fluxes from IRDIS are shown in Fig.\ref{fig:IFSspecsubstellar}.

For HD~151805, there are two candidate companions.
For HD~151805 CC1, we were unable to use MCMC sampling to determine the IRDIS $K_1$- and $K_2$-band flux, likely due to the companion being too faint. Instead, we relied on the flux estimate obtained from the second round of the NEGFC technique, after fixing the position during the first round. The errors were estimated by injecting a fake companion with the same flux and radial separation as CC1 at ten different position angles and taking the standard deviation of the measured fluxes. For the least squares fit, we left out the $K_2$-flux because S/N$_{K_2} < 2$. We find a mass of $0.03^{+0.01}_{-0.01}$ \Msun\ for CC1 and $0.04^{+0.02}_{-0.02}$ \Msun\ for CC2.

For CEN~16 CC2, we find a mass of $0.07^{+0.01}_{-0.02}$ \Msun\ from the second epoch observation. As mentioned earlier, the quality of the first epoch is not good enough to detect the candidate companion, which is clearly demonstrated in the PCA/SDI IFS image where CC2 is not visible (Fig.\ref{fig:SDIsubstellar}, panel b). There is another source present in the IFS image of CEN16 (CC1 in Fig.\ref{fig:SDIsubstellar}, panel b and c). However, since it is a bright source and not a low-mass companion, we do not discuss it here.

\section{Proper motion}
The images we retrieve from SPHERE represent a two-dimensional projection of the field of view around the central star. There is a priori no information on the distance to each of the detected sources, so they could also be background (or foreground) sources. Usually, true sources can be distinguished from background contaminants by taking a second epoch observation and testing for common proper motion with the central star. Co-moving sources are then identified as cluster members, attributing them a high probability of being bound to their central star, given that the density of the cluster is not too high \citep[see e.g.,][]{2021Janson}. A critical requirement for this technique to work is that the proper motion of the central star should be sufficiently different from that of the background contaminants.

\begin{figure}[!ht]
    \gridline{\fig{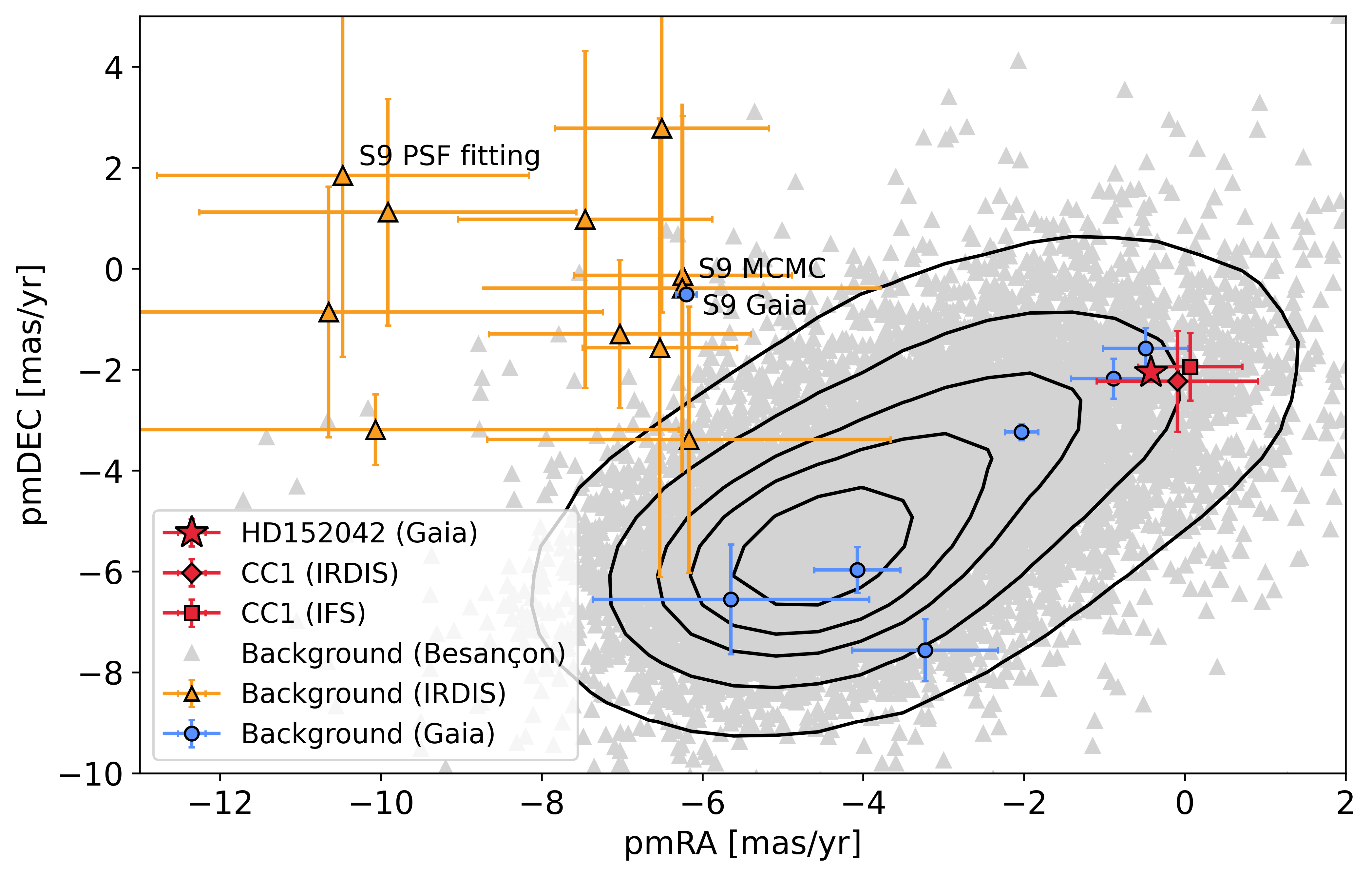}{0.47\textwidth}{}}
    \gridline{\fig{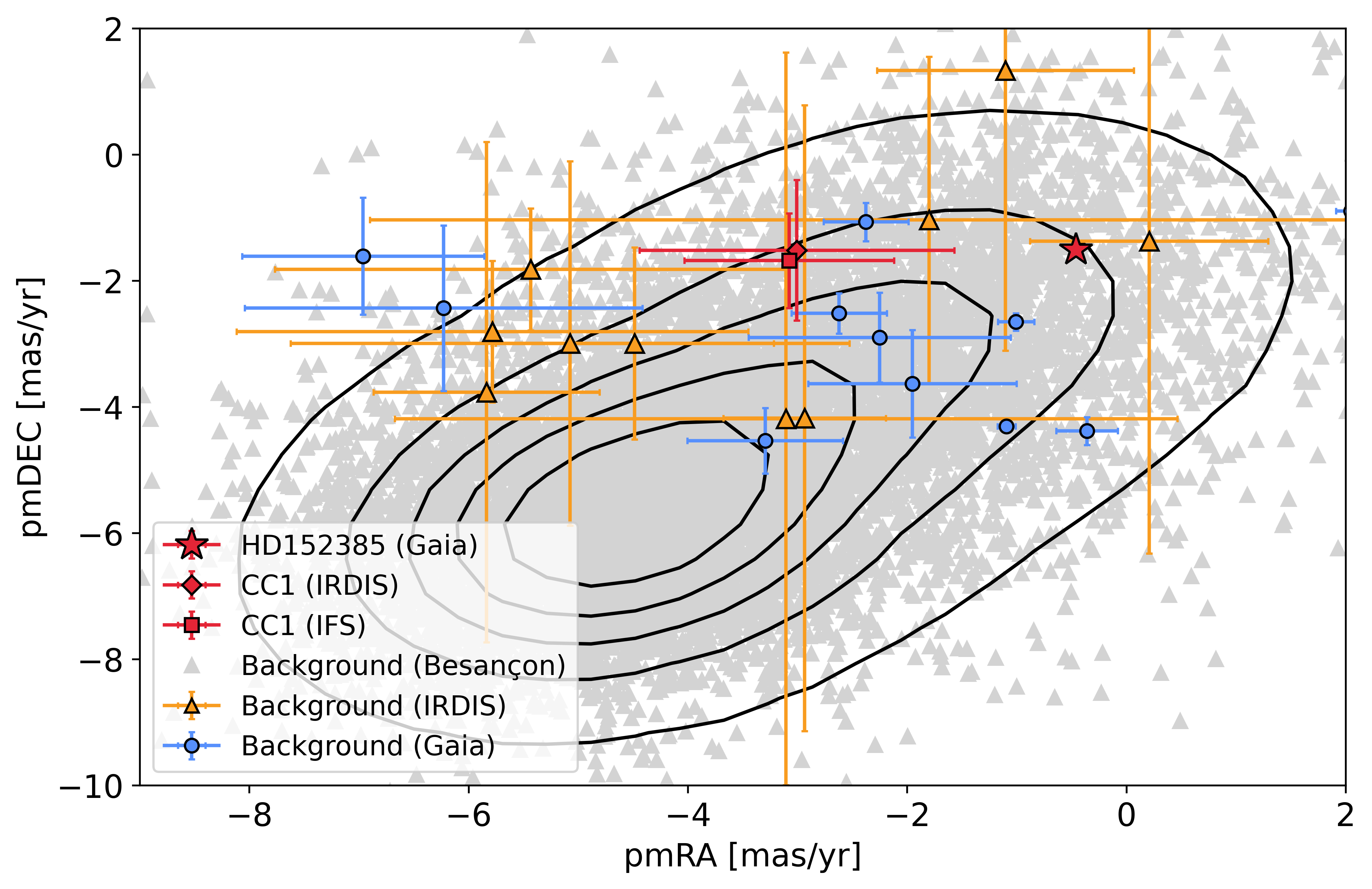}{0.47\textwidth}{}}
    \gridline{\fig{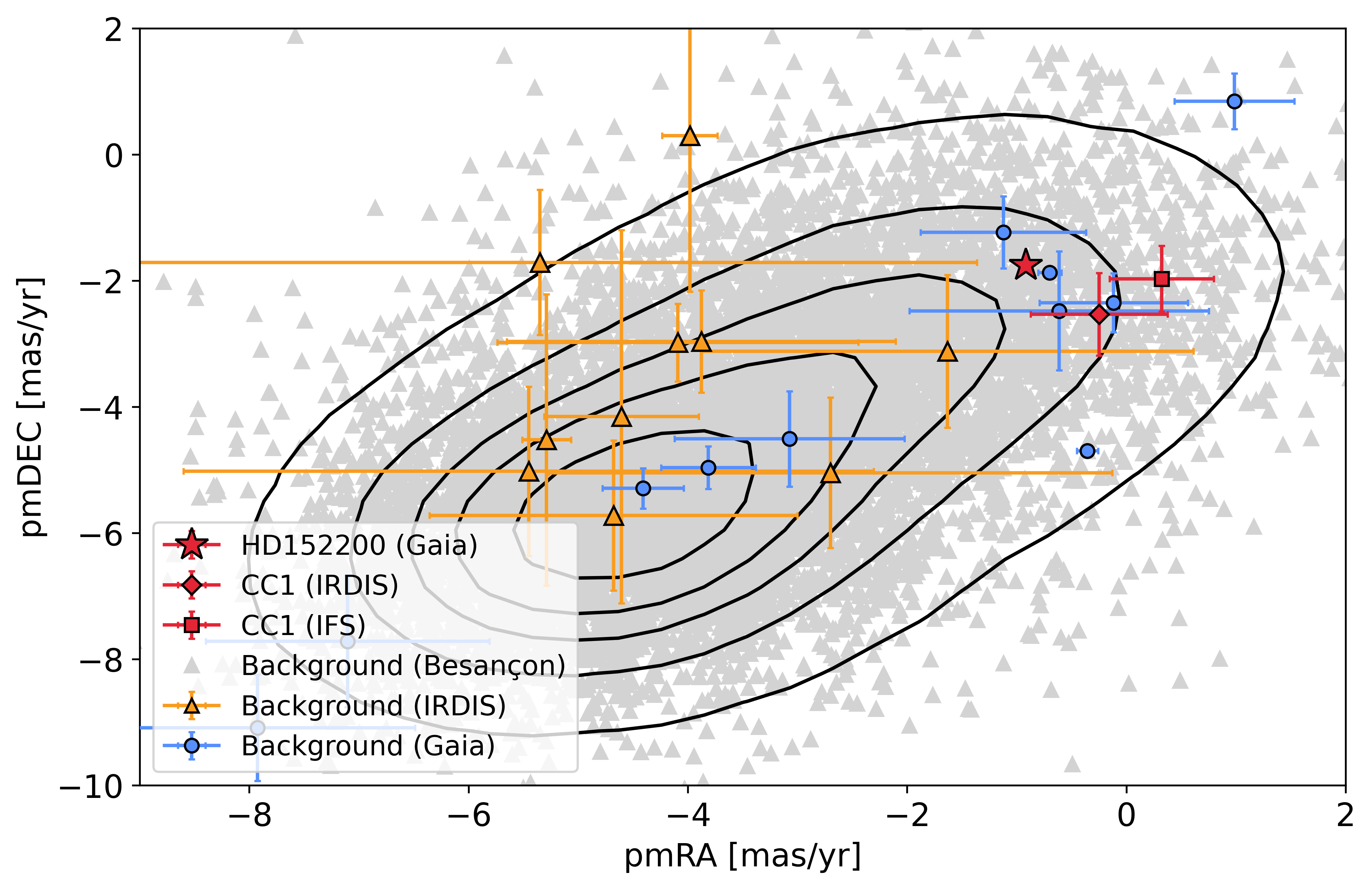}{0.47\textwidth}{}}
\caption{Proper motion of background sources from \citep[Besançon model of the Galaxy,][]{2022Robin}, Gaia \citep{2016Gaia,2023Gaia} and SPHERE/IRDIS for HD~152042, HD~152200 and HD~152385 compared to the proper motion of HD~152042, HD~152200 and HD~152385 \citep{2023Gaia, 2016Gaia} and their candidate companions. The contour levels are 0.05 (outer contour), 0.2, 0.4, 0.6 and 0.8 (inner contour).}
\label{fig:propermotion}
\end{figure}

\begin{table*}[!ht]
    \centering
    \caption{xy positions of candidate companions with two epochs.}
    \begin{tabular}{lccccc}
    \hline
    \hline
         Target & Date & $x_\mathrm{IFS}$ & $y_\mathrm{IFS}$ & $x_\mathrm{IRDIS}$ &$y_\mathrm{IRDIS}$ \\
         & & [pixels] & [pixels] & [pixels] & [pixels]  \\
         \hline
         HD~152042 CC1 & 2019-06-29 & $-66.14 \pm 0.24$ & $-20.54 \pm 0.22$ & $-40.52 \pm 0.22$ & $-12.83 \pm 0.22$ \\
         HD~152042 CC1 & 2023-05-14 & $-66.39 \pm 0.23$ & $-20.48 \pm 0.22$ & $-40.63 \pm 0.23$ & $-12.88 \pm 0.22$\\
         HD~152200 CC1 & 2015-08-19 & $-47.33 \pm 0.38$ & $-84.99 \pm 0.46$ & $-29.33 \pm 0.28$ & $-51.81 \pm 0.29$\\
         HD~152200 CC1 & 2023-05-16 & $-48.62 \pm 0.25$ & $-85.22 \pm 0.27$ & $-29.76 \pm 0.28$ & $-52.31 \pm 0.30$ \\ 
         HD~152385 CC1 & 2019-05-26 & $47.86 \pm 0.38$ & $-15.13 \pm 0.28$ & $28.96 \pm 0.32$ & $-9.36 \pm 0.25$\\ 
         HD~152385 CC1 & 2023-05-16 & $49.25 \pm 0.33$ & $-15.22 \pm 0.26$ & $29.79 \pm 0.34$ & $-9.36 \pm 0.26$ \\ 
         \hline
    \end{tabular}
    \label{tab:xypositions}
\end{table*}

\begin{table*}[!ht]
    \centering
    \caption{Proper motion of candidate companions}
    \begin{tabular}{lcccc}
    \hline
    \hline
         Target & $\mathrm{pmRA}_\mathrm{IFS,rel}$ & $\mathrm{pmDEC}_\mathrm{IFS,rel}$ & $\mathrm{pmRA}_\mathrm{IRDIS,rel}$ &$\mathrm{pmDEC}_\mathrm{IRDIS,rel}$ \\
         & [mas/yr] & [mas/yr] & [mas/yr] & [mas/yr]  \\
         \hline
         HD~152042 CC1  & $0.49 \pm 0.65$ & $0.11 \pm 0.67$ & $0.33 \pm 1.00$ & $-0.18 \pm 1.00$ \\
         HD~152200 CC1 &  $1.24 \pm 0.47$ & $-0.22 \pm 0.52$ & $0.67 \pm 0.62$ & $-0.78 \pm 0.65$ \\
         HD~152385 CC1 & $-2.61 \pm 0.95$ & $-0.17 \pm 0.75$ & $-2.54 \pm 1.44$ & $-0.01 \pm 1.11$ \\ 
         \hline
    \end{tabular}
    \label{tab:propmotion}
\end{table*}

\subsection{Proper motion of candidate companions}
We obtained second epoch observations for three out of the six targets in our sample (HD152042, HD152200 and HD152385). To compute the proper motion uncertainties we differentiated between `statistical' uncertainties (MCMC error, centering, True North, dither) that need to be calculated for every epoch and `systematic' uncertainties (platescale (including optical distortion), pupil offset, IFS offset) that can be added to the difference in position between the two epochs instead. The $x$ and $y$ positions derived from IFS and IRDIS observations including only the statistical uncertainties are presented in Table \ref{tab:xypositions}. The relative proper motion with respect to the central star in right ascension ($\mathrm{pmRA}_\mathrm{IFS,rel}$ and $\mathrm{pmRA}_\mathrm{IRDIS,rel}$) and declination ($\mathrm{pmDEC}_\mathrm{IFS,rel}$ and $\mathrm{pmDEC}_\mathrm{IRDIS,rel}$) direction is shown in Table \ref{tab:propmotion}, including both the statistical and systematic uncertainties. 

We find that there are inconsistencies between the positions measured from the IFS and IRDIS observations (Table \ref{tab:astrometry}). However, when looking at the proper motion (Table \ref{tab:propmotion}), the discrepancies disappear for HD~152042 and HD~152385. This suggests that they are likely due to systematic errors, such as calibration issues related to the platescale or orientation differences between the IFS and IRDIS instruments. Since these systematic effects are consistently observed in both epochs, they cancel out and do not impact our overall conclusions.

For HD~152200, the discrepancies seem to persist even in the proper motion measurements, especially in right ascension direction. However, calculating the difference between pmRA$_\mathrm{IFS,rel}$ and pmRA$_\mathrm{IRDIS,rel}$, we find $0.57 \pm 0.78$ mas/yr, which is not statistically significant. In the following, we adopt the weighted average of the IFS and IRDIS proper motion measurements. These are pmRA$_\mathrm{rel} = 1.03 \pm 0.37$ mas/yr and  pmDEC$_\mathrm{rel} = -0.44 \pm 0.41$ mas/yr.

\subsection{Proper motion of background sources}
Figure \ref{fig:propermotion} illustrates the proper motion of the candidate companions in comparison to the proper motion of the central star and various populations of background sources. The three panels correspond to different targets: HD152042, HD152200, and HD~152385. The proper motion data for the central stars were obtained from Gaia DR3 \citep[Table \ref{tab:targetchar},][]{2016Gaia, 2023Gaia}. To determine the proper motion of the candidate companions, we added the relative proper motion measured from IFS and IRDIS to the central star's proper motion from Gaia.
We conducted a simulation using the Besançon model of the Galaxy \citep{2022Robin} to estimate the proper motion of background stars within a solid angle of $0.1$ deg$^2$ centered around each target and included sources with $K\leq20$. From this simulation, we randomly selected $10000$ sources to plot in each panel. The contours, calculated from the Besançon simulation, represent cumulative frequency levels of $0.05$ (outer contour), $0.2$, $0.4$, $0.6$, and $0.8$ (inner contour).
Additionally, the figures display the proper motion of sources observed with SPHERE/IRDIS. We selected sources located beyond $2$\arcsec{} from the central star and calculated their positions at both epochs using a fast PSF fitting routine based on the python package \texttt{photutils} \citep{2020Rainot, 2020Bodensteiner,2019Bradley}, as the influence of the central star is negligible at separations of $\gtrsim 2$\arcsec{}. We added the proper motion of the central star to the measured relative proper motion of each IRDIS source. To maintain readability, we plotted the ten sources with the smallest error bars.
Finally, we included the proper motion of background sources from Gaia DR3 \citep{2023Gaia, 2016Gaia} within a 20\arcsec{} radius around each target. We excluded sources with a RUWE $\ge 1.4$ and those with a parallax between 0.5 and 0.8 mas to avoid plotting cluster members of Sco OB1 and potential companions of the targets. 

\subsection{Detailed examination of each target}
\subsubsection{HD~152042}
We find that the relative proper motion of the candidate companions around HD~152042 is consistent with zero within 1$\sigma$ (Table \ref{tab:astrometry}). However, Fig.\ref{fig:propermotion} shows that the proper motion of the IRDIS background sources does not align with the expectations derived from the Besançon simulation. Given that the Gaia measurements are consistent with the Besançon model, it is likely that the IRDIS measurements are biased, for example due to a residual effect from the centering of the IRDIS images or the PSF fitting method used. To investigate this, we remeasured the position of source `S9' \citep{2023Pauwels}, which is found in both the IRDIS data and Gaia DR3, with the VIP MCMC sampling method in two epochs. We found that the proper motion calculated from these remeasured positions matches the Gaia measurement, suggesting that the PSF fitting may be inaccurate for some sources.
Due to the extensive computation time required, we did not remeasure the positions of other sources with MCMC sampling. The critical point is that the proper motion of CC1, calculated with MCMC sampling, can be trusted. We conclude that the candidate companion is co-moving with the central star.

\subsubsection{HD~152200}
The relative proper motion of the candidate companion in RA direction differs by almost $3\sigma$ from zero. This difference could be partially explained by orbital motion, which in the case of a circular orbit is estimated to be $\lesssim 0.5$ mas/yr. With a time gap of $\sim 7.7$ years between the first and second epoch, the orbital motion might have been detectable depending on the inclination and eccentricity. Therefore, we do not reject it as a potential companion. In addition, although the error bars are large, many of the IRDIS sources seem to have substantially different proper motion from CC1.

\subsubsection{HD~152385}
The relative proper motion of the candidate companion in RA direction deviates from zero by almost $3\sigma$. However, the uncertainties are too large to draw a meaningful conclusion. Therefore, we cannot state that the source is co-moving at the moment. More precise measurements are required to reach a definitive determination.

\section{Spurious association probability}
Since there are background (and foreground) sources present that are co-moving with the targets, the fact that a source is co-moving does not provide a definitive confirmation that it is bound. Therefore, we use the magnitude and proper motion of the candidate companions to estimate the probability of spurious association, which is the probability that a source is not bound to the central star but is instead a result of chance alignment.

The calculation of the spurious association probability requires comprehensive knowledge of the magnitude and proper motion of the background contaminants down to sufficiently faint $K$-band magnitudes. Given that catalogues like VVV \citep{2010Minniti, 2010Saito} and 2MASS \citep{2006Skrutskie} do not extend to the faint magnitudes required and lack proper motion measurements, and considering that Gaia does not measure $K$-band magnitudes, we rely on the Besançon model of the Galaxy \citep{2022Robin} to represent the background source population. \cite{2023Pauwels} shows that the number of sources per magnitude bin calculated from the Besançon model indeed agrees very well with expectations from observed catalogues. For each target star, we adopt the same Besançon simulations as used in Fig.\ref{fig:propermotion}.

We computed the spurious association probability for a candidate companion with separation $r_i$ and magnitude $K_i$ using a Monte Carlo approach as described in \cite{2021Reggiani}, \cite{2022Rainot} and \cite{2023Pauwels}, and took into account background sources from the Besançon model of the Galaxy that apart from having a $K \leq K_i$ also meet the following criteria: 
\begin{eqnarray}
    \|\text{pmRA}_{\text{B}} - \text{pmRA}_{*}\| \leq \|\text{pmRA}_{\text{IFS,rel}}\| \nonumber 
    + \text{pmRA}_{\text{IFS,rel,err}} 
\end{eqnarray}
and
\begin{eqnarray} 
    \|\text{pmDEC}_{\text{B}} - \text{pmDEC}_{*}\| \leq \|\text{pmDEC}_{\text{IFS,rel}}\| \nonumber 
    + \text{pmDEC}_{\text{IFS,rel,err}}
\end{eqnarray}
with the proper motion in right ascension (RA) and declination (DEC) direction given by $\text{pmRA}_{*}$ and $\text{pmDEC}_{*}$ for the central star from Gaia \citep{2023Gaia,2016Gaia} and $\text{pmRA}_{\text{B}}$ and $\text{pmDEC}_{\text{B}}$ for the background sources from the Besançon model of the Galaxy. $\text{pmRA}_{\text{IFS,rel}}$ and $\text{pmDEC}_{\text{IFS,rel}}$ are the relative proper motion in both directions for the candidate companion (Table \ref{tab:propmotion}). We then generate 100000 times a population of background sources uniformally distributed over the FoV of the Besançon simulation (0.1 deg$^2$). The spurious association probability is given by the fraction of these populations for which at least one source is found at $r \leq r_i$.

Based on the IFS proper motion measurements (which are more precise than the IRDIS measurements), we find spurious association probabilities of $8 \cdot 10^{-4}$ for HD~152042 CC1 and $9 \cdot 10^{-3}$ for HD~152200 CC1. For HD~152385 CC1, we do not include the spurious association probability, as we do not consider it co-moving. Compared to the spurious association probabilities in \cite{2023Pauwels} (0.01 for HD~152042 CC1 and 0.08 for HD~152200 CC1), which do not include the proper motion information, this is factor $\sim 10$ improvement.

For the candidate companions for which we do not have a second epoch observation, we calculated the spurious association probabilities without including the proper motion constraint. The spurious association probabilities are 0.003 for CPD~-41\degree 7721 CC1 \citep{2023Pauwels}, 0.17 for HD~151805 CC1, 0.09 for HD~151805 CC2 and 0.05 for CEN~16 CC2.

\section{Discussion and conclusion}
\begin{figure*}
\gridline{\rightfig{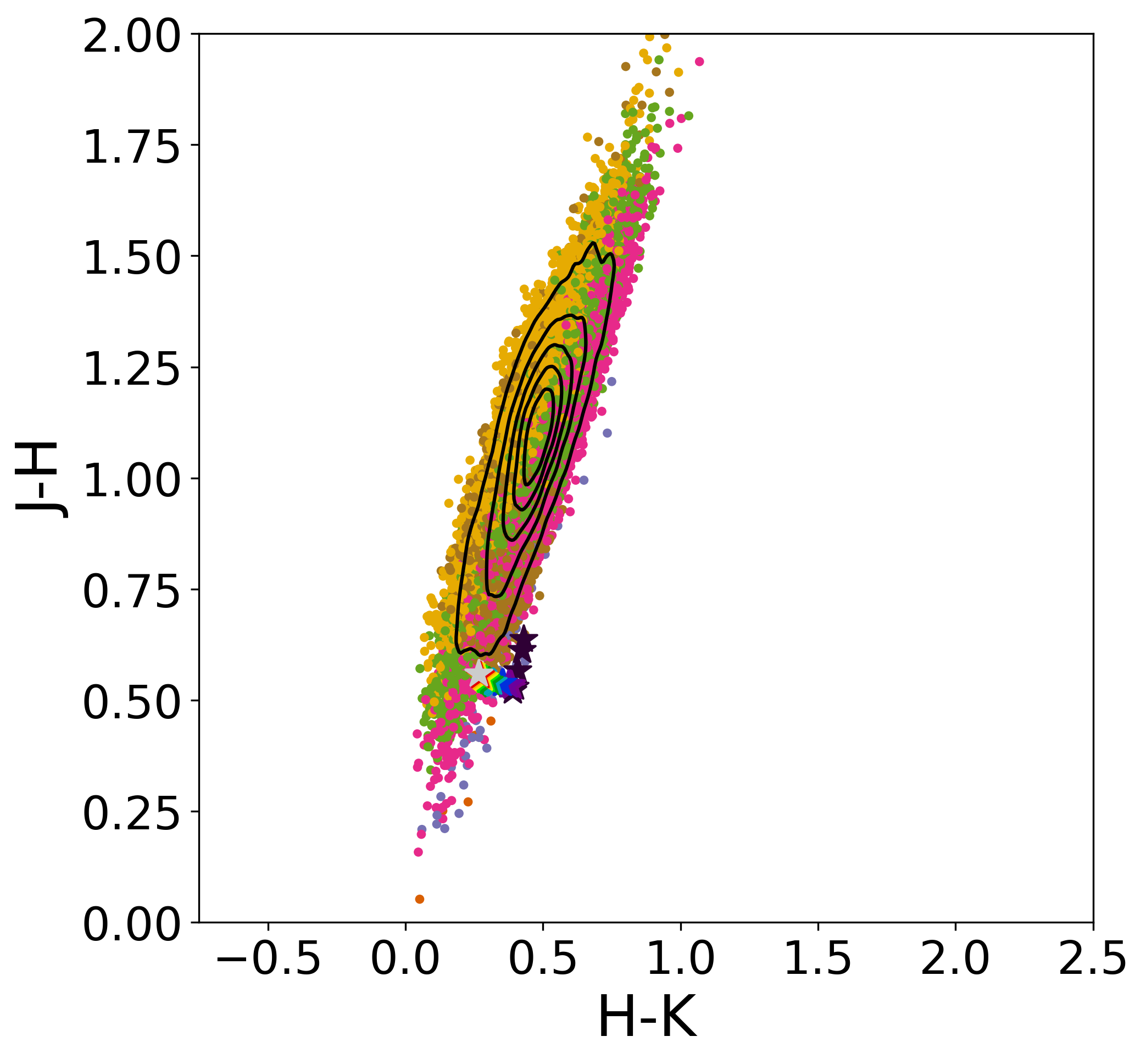}{0.28\textwidth}{(a)}
\rightfig{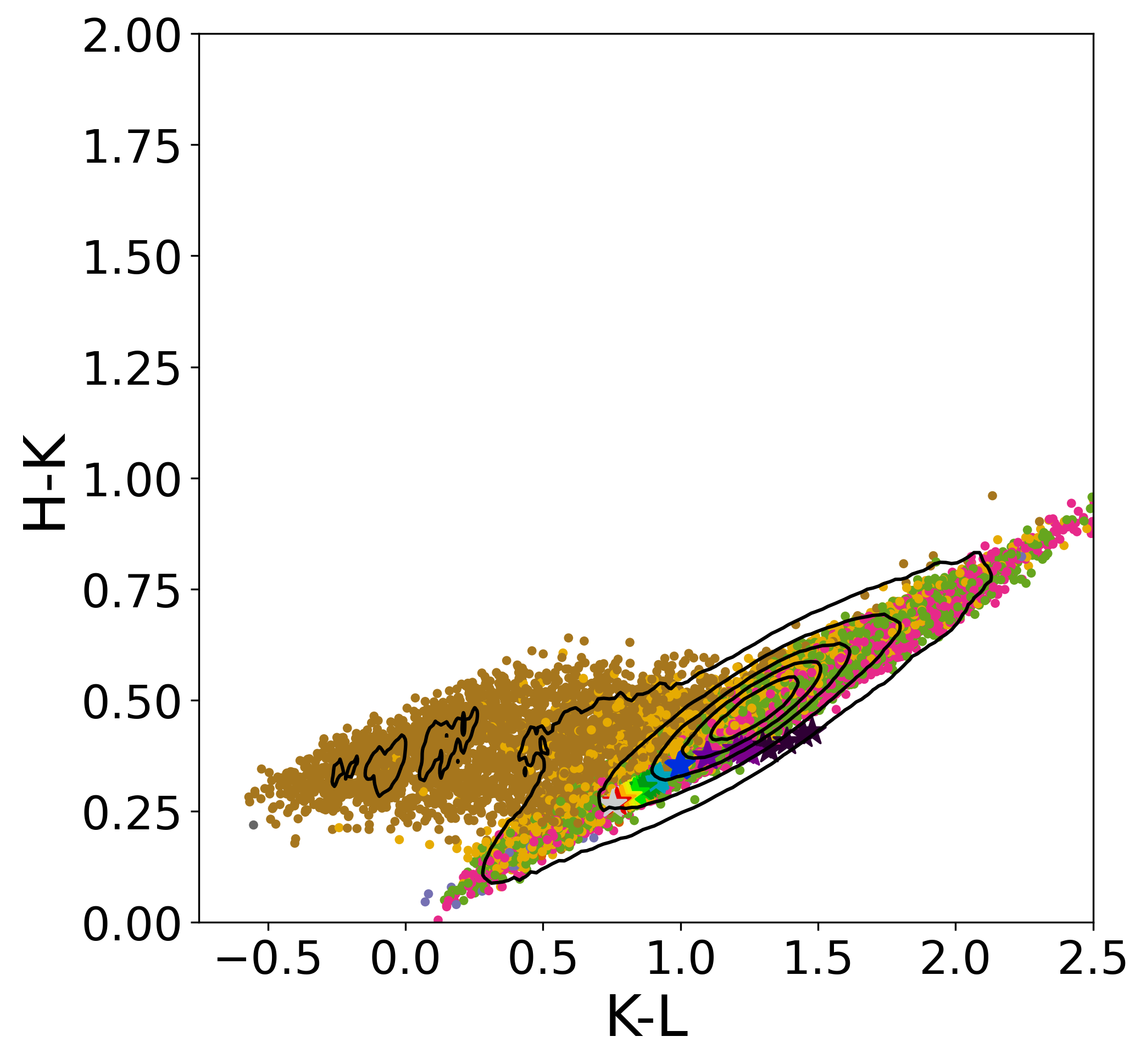}{0.28\textwidth}{(b)}
\rightfig{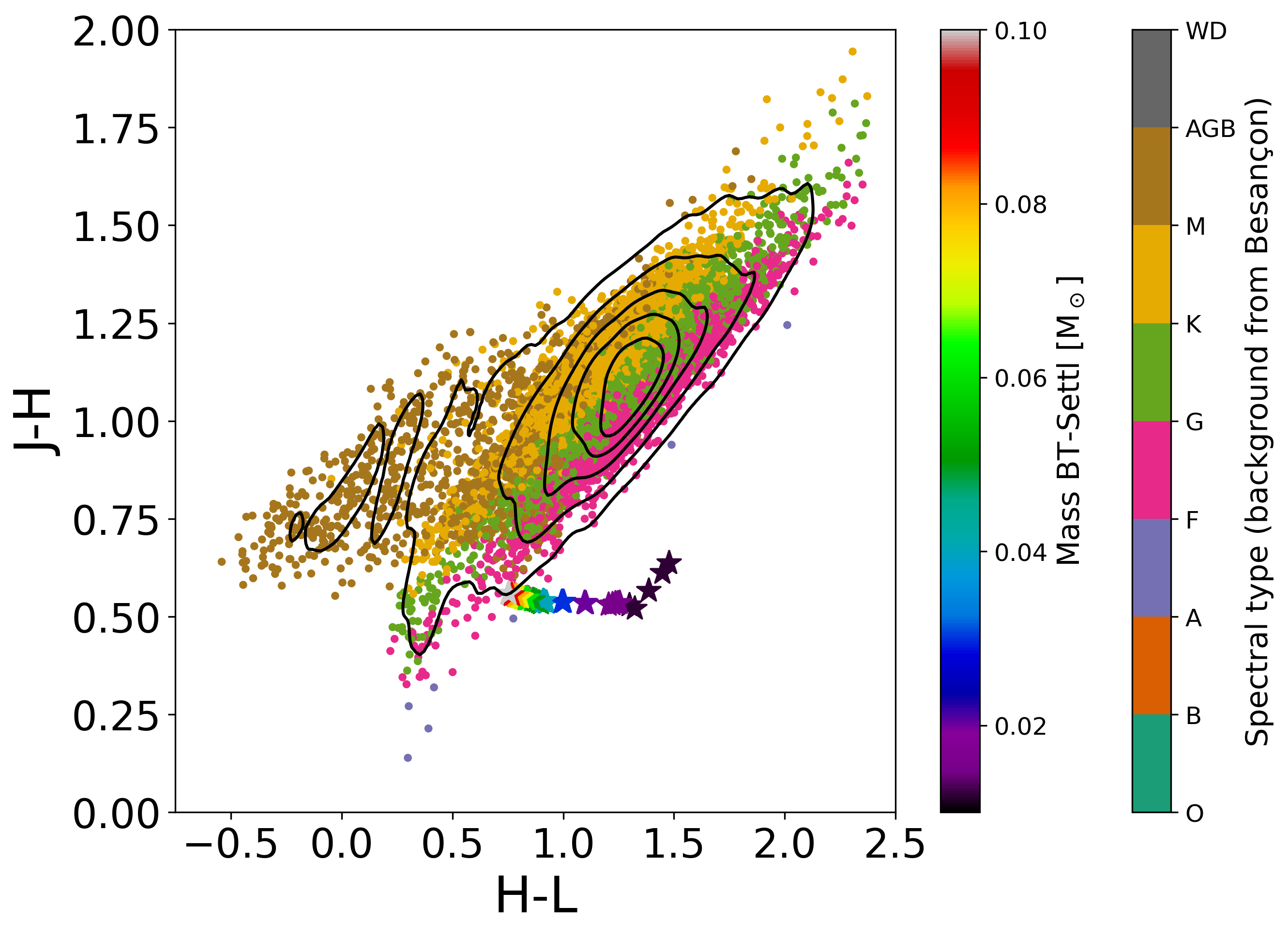}{0.36\textwidth}{(c)}}

\gridline{\fig{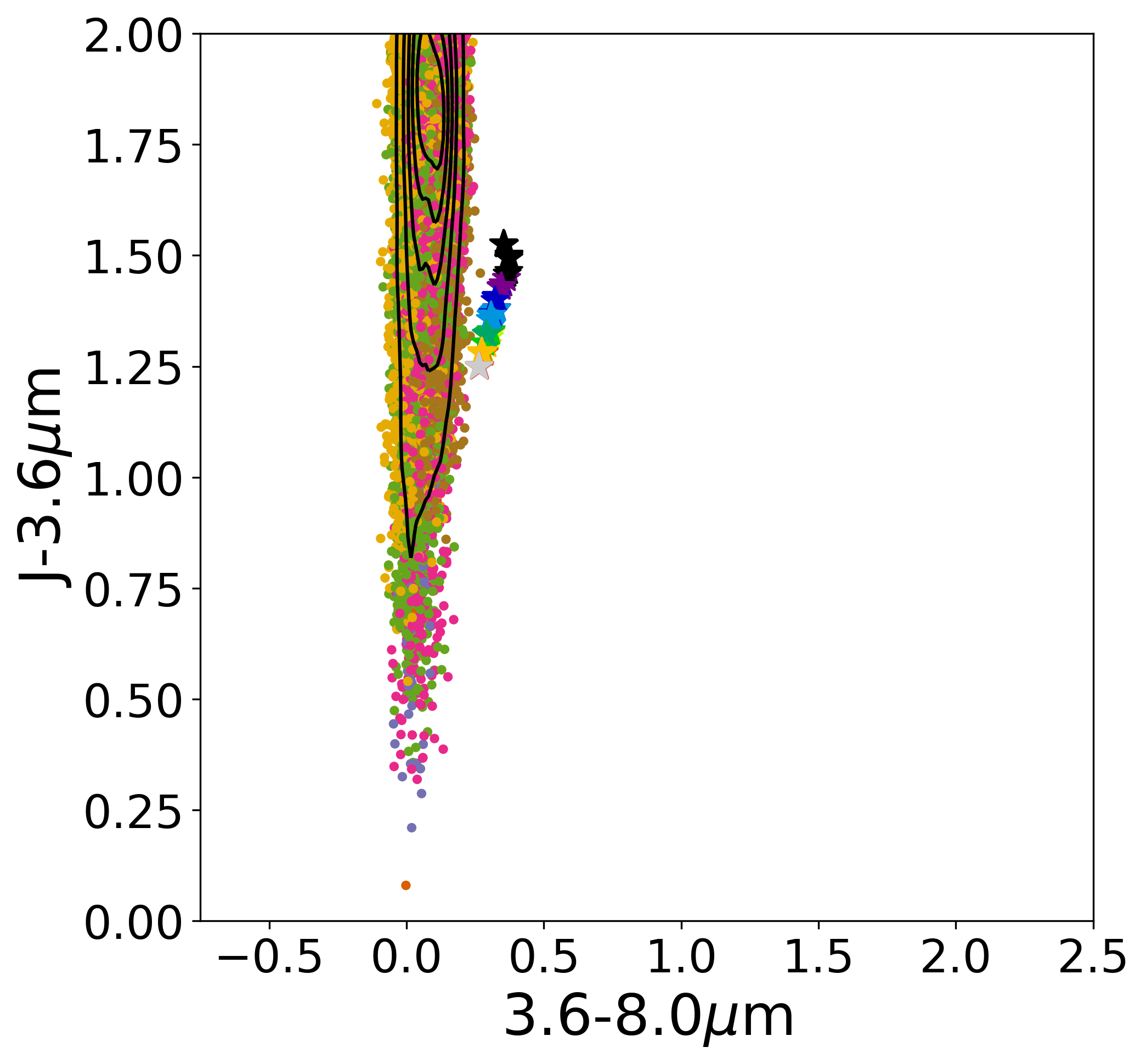}{0.28\textwidth}{(d)}
          \fig{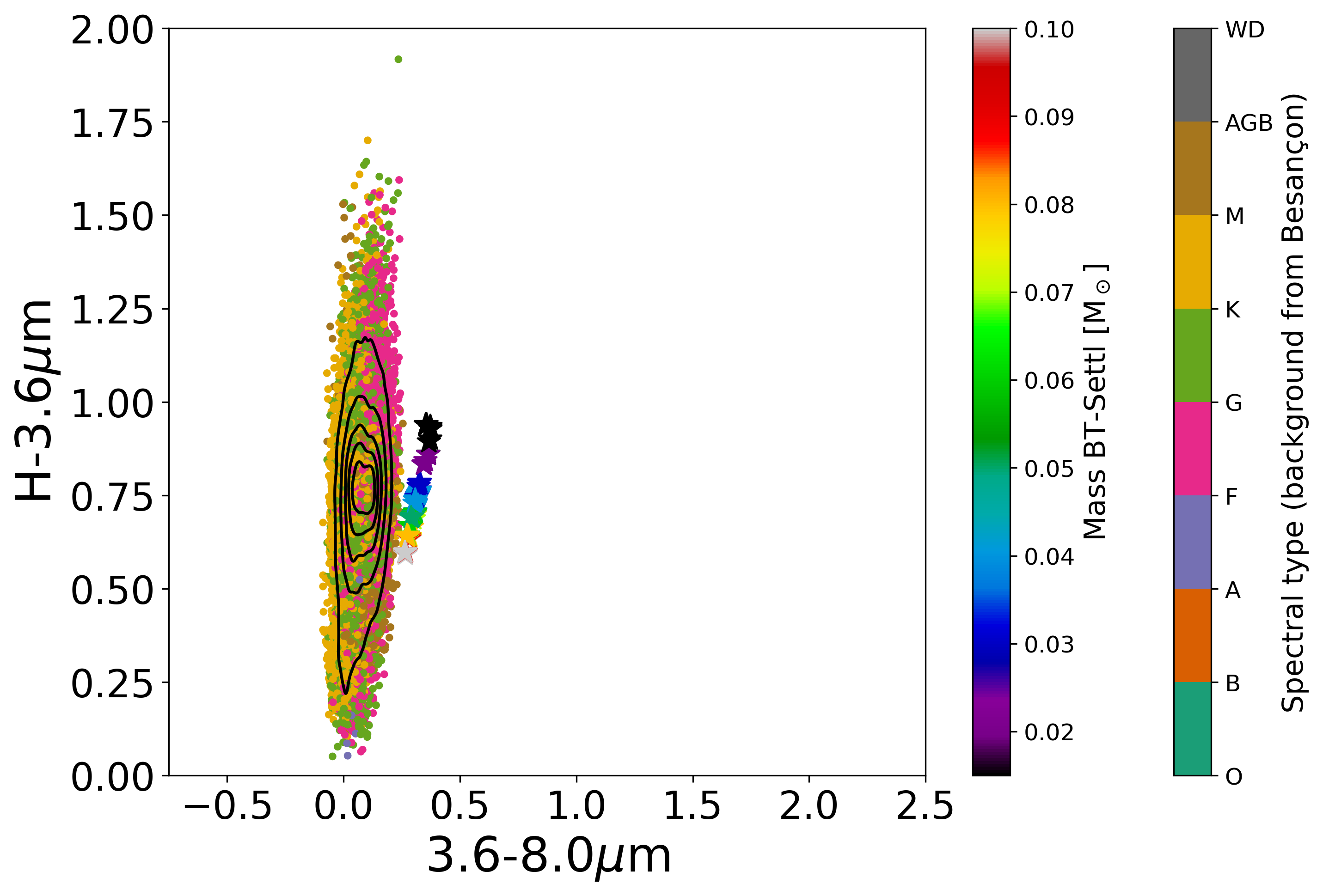}{0.385\textwidth}{(e)}}
\caption{$JHK$ (a), $HKL$ (b), $JHL$ (c), $J-3.6 \mu \text{m} - 8.0 \mu \text{m}$ (d) and $H-3.6 \mu \text{m} - 8.0 \mu \text{m}$ (e) color-color diagrams of a simulated population of background objects \citep[Besançon model of the Galaxy,][]{2022Robin} in the line of sight of Sco OB1. The expected colors of young brown dwarfs ($\leq$ 10 Myr) are computed from the BT-Settl models. The figures are plotted in the same scale for comparison. The black lines represent contours with the same levels as Fig.\ref{fig:propermotion}.}
  \label{fig:CCDs}
\end{figure*}
\subsection{Brown dwarf confirmation: future prospect}
Since the central stars are co-moving with the background sources, a distinction cannot be made based on their proper motion alone. Especially for the brown dwarf candidate companions, for which the spurious association probability is relatively high due to high background contamination at faint magnitudes, confirmation through an alternative method is desirable. Given the strong temperature and reddening difference of brown dwarfs compared to background FGKM stars, one might expect a color-color diagram (CCD) to be a good tool. Fig.\ref{fig:CCDs} shows a simulated population of background objects from the Besançon model of the Galaxy in a solid angle of 0.1 deg$^2$ around HD~151805 with $K\leq 20$. We randomly selected 50000 sources of the simulation to plot. In addition, the figure displays the expected colors of brown dwarfs from the BT-Settl models \citep{2014Allard}. We find that brown dwarfs can easily be separated from background FGKM stars in a  $J-H$ vs $H-L$ CCD if a precision of $\sim 0.01-0.1$ mag is reached. A similar CCD built only with $JHK$-measurements or $HKL$-measurements does not offer the needed diagnostic capability. This implies that observations covering the entire $J$- to $L$-wavelength range are essential to obtain a definite confirmation on the candidate companions' nature. These observations could be obtained by a combination of VLT/SPHERE (J and H band) and the Near Infrared Camera (NIRCam) instrument at the James Webb Space Telescope (JWST) (H and L band).

Similarly, a $J-3.6 \mu \text{m} - 8.0 \mu \text{m}$ or $H-3.6 \mu \text{m} - 8.0 \mu \text{m}$ CCD also allow us to distinguish brown dwarfs and background FGKM stars, although a better photometric precision is required compared to the $JHL$ CCD. The $3.6 \mu$m band could be obtained with JWST/NIRCam, although the inner working angle (IWA) at that wavelength is too large to observe the closest brown dwarf candidate companions at $\sim 0.5$\arcsec. There are currently no instruments that offer sufficiently deep observations in the $8.0 \mu$m band. A caveat is that brown dwarfs that are exposed to strong UV irradiation may exhibit peculiar colors due to photochemistry in their upper atmosphere, which are not accounted for by the BT-Settl models. Addressing this potential issue is beyond the scope of this paper.

\subsection{Formation pathways}
According to \cite{2022Offner}, binary formation mechanisms can be divided into four categories: filament fragmentation, core fragmentation, disk fragmentation and capture. The first three mechanisms consider in-situ formation of the multiple system, while capture requires dynamical interaction with other stars. We will consider the different formation theories in regard to the binary systems discussed in this paper. The primaries are mostly OB-type dwarfs and subgiants, with the exception of one B1Ib-type star, and have estimated masses between $9 - 17$ \Msun\ \citep{2017RamirezTannus, 2023Pauwels}. Their companions have masses $< 0.30$ \Msun.

Core and filament fragmentation occur when over-densities arise in the pre-stellar core or filament in which the core resides. These over-densities may collapse to form a (widely separated, $> 500$ AU) companion. Since the fragmentation of the core/filament occurs while there is still a large gas reservoir available in its surroundings, both stars will be able to accrete from this reservoir, making the formation of low mass ratio systems unlikely \citep{2022Offner}. 

Disks that are gravitationally unstable are prone to fragmentation and may form companions in this way \citep{2006Kratter}. A critical factor is the disk dispersal timescale. If the disk lives long enough and fragments before most of the mass has been accreted by the central star, the companion may also accrete from the disk, producing more equal mass binaries \citep{2010Kratter}. However, disks around early-type stars are expected to disperse rapidly \citep[][and references therein]{2023Damian}, favouring systems with low mass ratios. 
\cite{2020Oliva} performed self-gravity-radiation-hydrodynamics simulations of disk fragmentation around a massive star, showing that fragments typically have masses $\sim 1$ \Msun, which is significantly higher than the low-mass companions discussed in this paper. These companions are therefore directly challenging binary formation mechanisms, since none of the in-situ formation scenarios predict such low-mass companions to be formed around massive stars.
However, the simulations are limited to timescales of the order of tens of kyr because of their extensive computation time. Therefore, the further evolution of such systems is unknown: e.g., fragments may interact with each other and with the disk, or merge with the central star. 
In terms of observations, a massive star with a fragmenting disk containing a low-mass fragment has been detected by the Atacama Large Millimeter/submillimeter Array (ALMA). \cite{2018Ilee} observed a $\sim 40$ \Msun\ proto-O star with a $< 0.6$ \Msun\ fragment at 1920 AU. Such a system could potentially be a progenitor to the low-mass stellar companions discussed in this paper, although it contains an even higher mass primary star.

Finally, capture may play an important role in the formation of low mass ratio binary systems. A priori, this mechanism cannot be distinguished from in-situ formation mechanisms. Instead, one has to rely on statistical considerations to identify the likelihood that a low-mass companion is captured by the massive star. This depends, among others things, on the stellar density and the age of the cluster, the mass of the central star, and the mass of the companion. \cite{2022Daffern-Powell} perform N-body simulations of young star-forming regions and track their evolution over a period of 10 Myrs with the aim of identifying how many planets are stolen (from another star) or captured (as free-floating object). They discover that captured planets generally have wider orbits than stolen or preserved planets and that the eccentricity and inclination distributions of captured and stolen planets are similar, but different from the eccentricity and inclination distribution of preserved planets. Although the simulations should be repeated taking into account massive primaries and companions with higher masses (brown dwarfs and low-mass stellar companions), such a result can directly be compared with the observed number of low-mass companions around massive stars. Should one detect many more companions than predicted by capture, this strongly argues that another formation mechanism is also at play. Measuring the eccentricity and inclination distributions would be valuable, but this is challenging because of the long orbital periods of our systems ($\sim 10^3$ yrs). 

In conclusion, the formation of these low mass ratio systems remains open for debate. It is unclear whether in-situ formation mechanisms are plausible, as this would imply that low-mass proto-stars and proto-brown dwarfs are able to survive in the direct UV-irradiated environment powered by the central massive star. In order to understand this, it is of crucial importance to confirm (or not) the brown dwarf and low-mass stellar companions through second epoch observations, establishing their colors and proper motion. Our observations show that at least one or two low-mass stellar companions share similar proper motion to their central star and have a high probability to be bound, offering unprecedented possibility of testing this scenario. However, it is necessary to increase the statistics on these low-mass (sub)stellar companions to investigate whether they can all be explained by capture or if another formation mechanism should be considered.

\section{Acknowledgements}
This work has made use of the High Contrast Data Center, jointly operated by OSUG/IPAG (Grenoble), PYTHEAS/LAM/CeSAM (Marseille), OCA/Lagrange (Nice), Observatoire de Paris/LESIA (Paris), and Observatoire de Lyon/CRAL, and supported by a grant from Labex OSUG@2020 (Investissements d’avenir – ANR10 LABX56).

Moreover, this work has made use of data from the European Space Agency (ESA) mission
{\it Gaia} (\url{https://www.cosmos.esa.int/gaia}), processed by the {\it Gaia}
Data Processing and Analysis Consortium (DPAC,
\url{https://www.cosmos.esa.int/web/gaia/dpac/consortium}). Funding for the DPAC
has been provided by national institutions, in particular the institutions
participating in the {\it Gaia} Multilateral Agreement.

Furthermore, we are grateful for the support received from the European Research Council (ERC) under the European Union’s Horizon 2020 research and innovation programme (grant agreement number 772225: MULTIPLES). 

Finally, T.P. thanks the Research Foundation - Flanders (FWO) for the PhD fellowship 1164522N.


\bibliography{sample631}{}
\bibliographystyle{aasjournal}



\end{document}